\documentclass[12pt,a4paper]{article}
\usepackage[a4paper,margin=2cm]{geometry}

\usepackage{graphicx}
\usepackage{amsmath}
\usepackage{mathtools}
\usepackage{amssymb}
\usepackage{grffile}
\usepackage{float}
\usepackage{xcolor}
\usepackage{gensymb}
\usepackage{hyperref}
\usepackage{dcolumn}
\usepackage{bm}
\usepackage{multirow}
\usepackage{makecell}
\usepackage{soul}
\usepackage{caption}
\usepackage{placeins}
\usepackage{bbm}
\usepackage{upgreek}
\usepackage{booktabs}
\usepackage{rsc}
\usepackage{hyperref}

\vspace{-1.5em}
\title{Understanding Graphene–Perovskite Interactions: From Flake Chemistry to Crystallisation and Solar Cell Performance }
	
\vspace{-1.0em}
\author{
	Oussama Er-Riyahi,\textit{$^{1,2,6}$}$^\diamond$ 
	Dorota Biernacka,\textit{$^{1,3}$}$^\diamond$ 
	Yating Guo,\textit{$^4$}$^\diamond$ \\
	Mikel Ballarena Tellechea,\textit{$^{1,2}$} 
	Silver-Hamill Turren-Cruz,\textit{$^5$} 
	Dario Bercioux,\textit{$^{6,7}$} \\
	Meng Li,\textit{$^4$}$^\ast$ 
	Agnieszka Lekawa-Raus,\textit{$^3$}$^\ast$ 
	Jorge Pascual\textit{$^{8,9}$}$^\ast$ \\
	Karolina Z. Milowska,\textit{$^{1,10,7}$}$^\ast$
}
	\date{}

\begin{document}

\newgeometry{top=1.1cm,bottom=1.6cm,left=2cm,right=2cm}
\maketitle

\vspace{-2.0em}

\begin{center}
	\footnotesize  
	 $^1$CIC nanoGUNE, Tolosa Hiribidea 76, E-20018 Donostia, Basque Country, Spain\\
	 $^2$Fisika Saila, Centro F\'isica Materiales CSIC-UPV/EHU, E-48080 Bilbo, Basque Country, Spain\\
	 $^3$Centre for Advanced Materials and Technologies (CEZAMAT), Warsaw University of Technology, ul. Poleczki 19, 02-822 Warsaw, Poland; E-mail: agnieszka.raus@pw.edu.pl\\
	 $^4$Key Laboratory for Special Functional Materials of Ministry of Education, School of Nanoscience and Materials Engineering, Henan University, 475004 Kaifeng, China;  E-mail: mengli@henu.edu.cn\\
	 $^5$Instituto de Ciencia de los Materiales (ICMUV), Universitat de Valencia, 46980 Paterna, Spain\\
	 $^6$Donostia International Physics Center, Paseo Manuel de Lardizabal 4, 20018 San Sebastian, Spain\\
	 $^7$Ikerbasque, Basque Foundation for Science, Plaza Euskadi 5, 48009 Bilbao, Spain\\
	 $^8$Instituto de Tecnología Química, Universitat Politècnica València-Consejo Superior de Investigaciones Científicas, Av. dels Tarongers, 46022 València, Spain; E-mail: jpasmie@itq.upv.es\\
	 $^9$Polymat, University of the Basque Country UPV/EHU, 20018 Donostia-San Sebastian, Spain\\
	 $^10$BCMaterials, Basque Center for Materials, Applications and Nanostructures, UPV/EHU Science Park, 48940 Leioa, Spain; E-mail: kz.milowska@bcmaterials.net\\
\end{center}
\begin{center}
	\footnotesize
	$^\diamond$ Contributed equally to this work.\\
\end{center}
\begin{center}
	\footnotesize
	$^{*}$ Corresponding authors: agnieszka.raus@pw.edu.pl; mengli@henu.edu.cn; jpasmie@itq.upv.es; kz.milowska@bcmaterials.net; 
\end{center}

\vspace{-2.0em}

\abstract{Graphene-derivatives are widely employed materials to improve bulk and interface properties of metal-halide perovskite devices. Yet the implications of their flake chemistry and interactions with the perovskite precursors remain unclear. Here, we show that pristine graphene flakes (GF) and more conventional graphene oxide flakes (GOF) are not interchangeable. Density functional theory calculations reveal that GOF interacts more strongly with the perovskite lattice but induces larger structural distortions, stronger interfacial polarisation, and localised gap states. In contrast, GF forms comparatively non-disruptive contacts, a response retained across a wide compositional range. Machine-learning atomistic simulations further show that GF contacts both Pb- and I-containing regions of solvated perovskite nanocrystals, with a strong solvent dependency. Solution spectroscopic characterization indicates that GF additives serve as scaffold for preorganised Pb/I-containing precursors, favouring film crystallisation. In this sense, GF enhances solar cell performance across perovskite compositions, but particularly those facing a more challenging crystallisation. In mixed Sn–Pb perovskite solar cells, GF raises the champion power-conversion efficiency from 21.5\% to 23.7\% with improved storage stability. These results establish pristine GF as a chemically defined additive and connect its atomic-scale interactions with precursor organisation, crystallisation, device performance, and stability.
}

\clearpage
\restoregeometry

\section{Introduction}
\color{black}
The wet processing of metal halide perovskites is highly sensitive to the coordination chemistry and colloidal organisation of their precursor species, which govern nucleation, crystallisation, film morphology, and ultimately device performance\cite{Yan2015,Radicchi2019}. Controlling solution environment is particularly important for materials that face limitations in their crystallization mechanism, such as Sn-containing perovskites, which show higher precursor reactivity and lower colloidal instability.~\cite{Pascual2026, Hu2025} Additive engineering has therefore become a widely used strategy to control crystallisation by modifying the coordination chemistry and colloidal organisation of the precursor solution through chemical functionalization~\cite{Maschwitz2025, Hu2025}. However, the functional groups introduced by these additives should improve the precursor environment without introducing detrimental defects into the resulting films.

Among the wide range of molecular families investigated as additives in perovskite precursor solutions, carbon nanomaterials have emerged as versatile scaffolds with the ability to design specific chemical interactions. In particular, graphene-based materials have been extensively incorporated into perovskite solar cells, but their effects depend strongly on their chemical structure and location within the device.~\cite{Gong2021, Milic2018}
Graphene oxide (GO) contains hydroxyl and epoxy groups across its basal plane and carbonyl and carboxyl groups mainly at the edges, while stabilising oxidative debris commonly forms an additional layer on its surfaces.~\cite{dreyer2010, Pei2012}
This oxygen-rich structure promotes dispersibility in polar precursor solutions but makes GO chemically heterogeneous and structurally ill-defined. Reduced graphene oxide (rGO) partially restores the conjugated sp$^2$ network, although residual oxygen groups and structural defects remain.
In contrast, pristine graphene is characterised by an unfunctionalised, extended sp$^2$-bonded carbon lattice. Commercial powders marketed as pristine graphene, however, are ensembles of a finite number of flakes and may differ substantially in layer-number and lateral-size distributions, morphology, defect density, and edge-to-basal-plane ratio.~\cite{Kauling2018,Alves2024}
Among these materials, pristine graphene flakes obtained by non-oxidative top-down exfoliation of graphite~\cite{Novoselov2004,Paton2014,Padi2024} or bottom-up plasma synthesis from methane~\cite{LevidianG3TDS} provide a scalable form of graphene suitable for dispersion-based processing.~\cite{backes2020production} 
With lateral dimensions extending up to approximately 1--2~$\mu$m, these one-to-few-layer flakes retain the semimetallic character associated with an intact sp$^2$ basal plane, while their finite edge-to-basal-plane ratio gives rise to amphiphilicity and improves their processability in polar media.~\cite{kuziel2020} Unlike GOF, the amphiphilicity of pristine GF arises from its finite-flake topology, which combines a hydrophobic basal plane with comparatively hydrophilic edges, without requiring oxygen functionalisation.
These distinctions are important because oxygen functionalisation can improve dispersibility while simultaneously altering the interaction with the perovskite.

Direct incorporation of graphene-based materials into the perovskite absorber has received considerably less attention than their use in charge-transport layers, electrodes, and interfacial contacts. Most absorber-additive studies have used GO, rGO, or chemically doped graphene derivatives in Pb-based perovskites.~\cite{Hadadian2016, Zhang2018GO, Balis2020, Kim2020rGO, Marchezi2021, Azzaz2025}
Only a few studies have examined pristine graphene materials directly within the perovskite-forming layer.~\cite{Redondo2021, Li2018GrapheneNanofibers} These reports demonstrate changes in nucleation, grain morphology, efficiency, and stability, but the underlying graphene--perovskite interaction remains largely inferred from the resulting films and devices. Graphene derivatives have also been introduced into pure-Sn absorbers, including rGO--Sn quantum-dot and N-doped-GO composites.~\cite{Mahmoudi2021SnGraphene, Mahmoudi2022NGO} However, the direct use of pristine few-layer graphene in mixed Sn--Pb precursor solutions and its influence on precursor organisation remain largely unexplored. 

Here, we combine density functional theory and machine-learning atomistic simulations with solution-phase spectroscopy, film characterisation, and photovoltaic measurements to determine how pristine GF interacts with halide perovskites across multiple length scales. We first compare pristine and oxidised graphene flakes at halide-perovskite surfaces and establish how flake chemistry controls structural distortion, electronic states, charge redistribution, and surface energetics. We then use solvated $\mathrm{CsPbI_3}$ films, UV--vis spectroscopy, and $^{207}$Pb NMR to show that GF interacts with Pb- and I-containing environments and promotes precursor pre-organisation in solution. This reorganisation improves crystallisation and film quality across compositions, particularly for narrow-bandgap mixed Sn–Pb perovskites, with device efficiency increasing from 21.5\% to 23.7\% while markedly improving stability. The results identify pristine GF as a chemically well-defined additive for controlling perovskite formation without the strongly perturbing structural and electronic effects associated with oxidised flakes.

\color{black}{}
\section{Results and discussion}

\begin{figure*}[!t]
    \centering
    \includegraphics[width=0.98\textwidth]{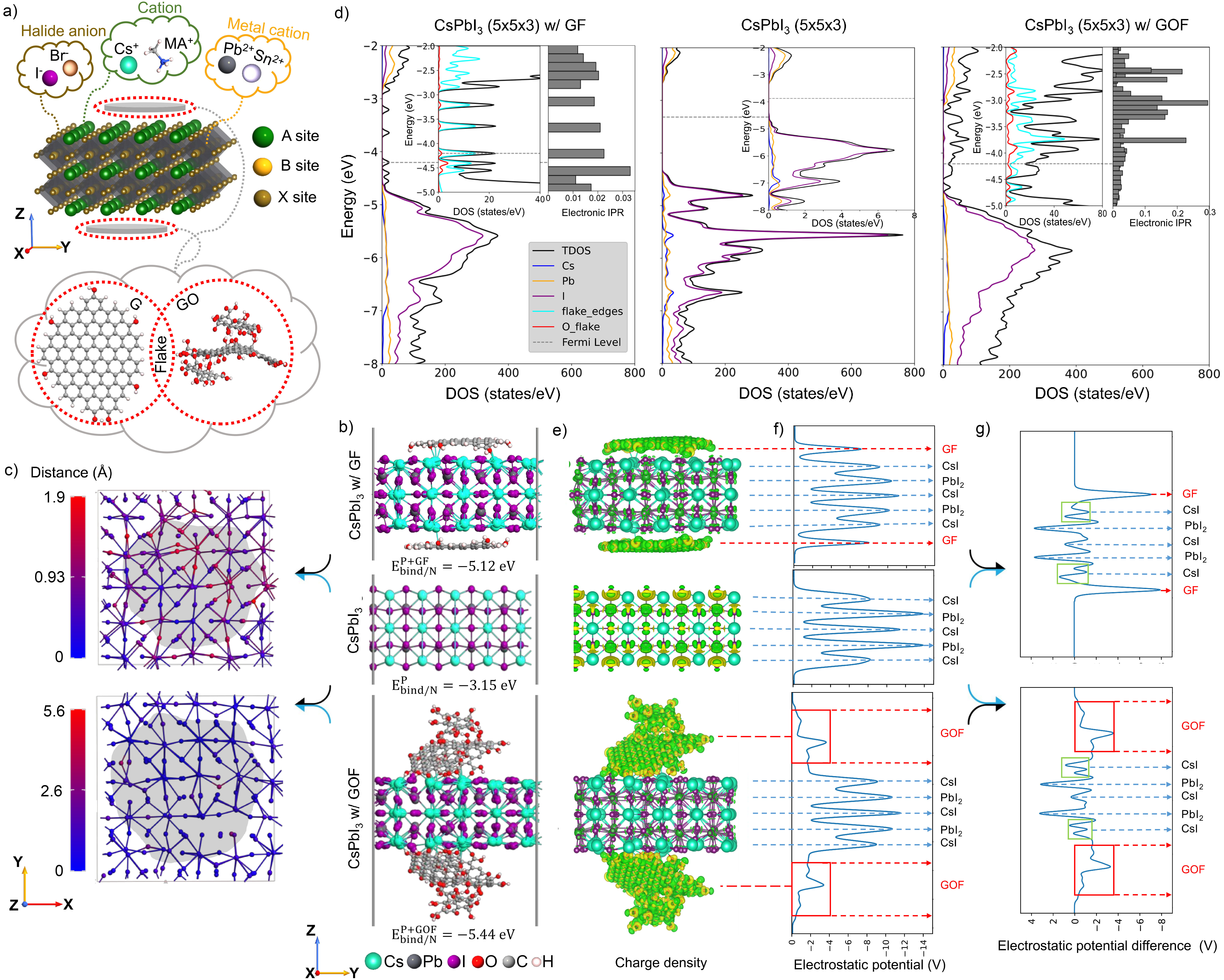}\\[0.6em]
    \caption{Contrasting atomic-scale interactions of graphene (G) and graphene oxide (GO) flakes with halide perovskites.
    a)Schematic illustration of the graphene-based/perovskite heterostructures studied at the DFT level, consisting of ligand-free $\mathrm{ABX_3}$ perovskite slabs in contact with G or GO flakes on both exposed surfaces. A represents the A-site cation (Cs$^{+}$ or MA$^{+}$), B the B-site metal cation (Pb$^{2+}$ or Sn$^{2+}$), and X the halide anion (I$^{-}$ or Br$^{-}$).
    b) Side views of the DFT/PBEsol-optimised three-layer $\mathrm{CsPbI_3}$ $(5\times5\times3)$ slab interacting with GF (top) or GOF (bottom), together with the non-interacting reference slab (middle). The corresponding binding energies per atom are indicated.
    c) Atomic displacement maps for the GF- and GOF-containing slabs relative to the pristine slab, showing the smaller perturbation induced by GF. The flakes are shown in grey, while the slab atoms are coloured from blue (0~\AA) to red according to their absolute displacement.
    d) Total and projected density of states (TDOS/PDOS) for the pristine $\mathrm{CsPbI_3}$ $(5 \times 5 \times 3)$ slab (middle), compared with the slab interacting with GF (left) and GOF (right). The DOS is projected onto Cs, Pb, and I atoms of the perovskite slab, as well as O atoms (all O in GOF) and edge C atoms of the G and main GO flakes. Insets in the left and right panels show enlarged views of the band-gap region together with the corresponding electronic inverse participation ratio (IPR), allowing identification of states introduced within the perovskite band gap by the flakes. The inset in the middle panel shows the density of states of cubic bulk $\mathrm{CsPbI_3}$, projected onto Cs, Pb, and I atoms, for comparison with the slab electronic structure.
    e)Charge-density difference isosurfaces for slab interacting with GF (top) and GOF (bottom), together with the pristine slab (middle), obtained as the difference between the self-consistent electron density and the superposition of neutral atomic densities. The isovalues are 0.005, 0.002, and 0.006 e\,\AA$^{-3}$.  Green and yellow regions indicate charge accumulation and depletion, respectively. 
    f) Macroscopically averaged electrostatic potentials along the $z$ direction for the same three systems.
    g) Corresponding electrostatic-potential differences obtained by subtracting the pristine-slab profile from those of the GF- and GOF-containing slabs.
    }
    \label{fig:dft}
\end{figure*}

To understand interactions between graphene flakes and halide perovskites, we investigated the structural and electronic properties of various halide-perovskite/graphene-flake heterostructures using density functional theory (DFT) calculations. As schematically shown in Figure~\ref{fig:dft}a, we used three-layer-thick ($\sim1$~nm), ligand-free perovskite slabs with graphene flakes (1.5--2.4~nm in lateral dimension) placed symmetrically on both exposed surfaces. These minimal models are relevant to flakes printed onto perovskite thin films or incorporated between perovskite grains, and allow us to distinguish the direct interaction of the flake basal plane and edges with the perovskite surface and deeper slab regions while keeping the first-principles calculations feasible. To determine whether the choice of carbon flake affects the perovskite counterpart, we modelled two different flakes: commonly used graphene oxide flake (GOF) and pristine graphene flake (GF). 
These models represent two contrasting flake chemistries: GF retains an extended sp$^2$--bonded, $\pi$--conjugated carbon network and therefore the characteristic electronic nature of graphene~\cite{kuziel2020}, whereas the strongly oxidised and structurally heterogeneous GOF does not preserve this electronic character.
In our GF model, a small number of oxygen-containing groups was introduced only at the flake edges to reflect the low but non-zero oxygen content of experimentally available pristine graphene flakes without disrupting the graphitic basal plane.
To test whether the graphene-flake response depends on perovskite composition, we considered three representative halide perovskites: all-inorganic $\mathrm{CsPbI_3}$,~\cite{Kovalenko2017} widely studied hybrid organic--inorganic $\mathrm{MAPbI_3}$,~\cite{Chen2019, Wu2017, Li2024} and lead-free all-inorganic  $\mathrm{CsSnBr_3}$.~\cite{Gupta2016, Kheswa2025, Islam2026} Together, these perovskites span organic and inorganic A-site cations, Pb- and Sn-based B-site cations, and iodide and bromide anions. The $\mathrm{CsPbI_3}$/GF and $\mathrm{CsPbI_3}$/GOF heterostructures were used for the direct comparison between pristine and oxidised graphene flakes (Figure~\ref{fig:dft}b), whereas $\mathrm{MAPbI_3}$ and $\mathrm{CsSnBr_3}$ were considered only in combination with GF (Figure~\ref{fig:s10}) to test whether the GF--induced response persists across different perovskite compositions.

The interaction between the $\mathrm{CsPbI_3}$ slab and the flakes leads to distinct structural responses depending on the flake chemistry (Figure~\ref{fig:dft}b,c). GF, which contains only a small number of oxygen-containing groups at the edges, induces limited distortion of the perovskite surface and largely preserves the slab geometry. In contrast, GOF produces larger and more spatially extended displacements of the perovskite atoms relative to the pristine slab, as shown by the displacement maps in Figure~\ref{fig:dft}c. This indicates a stronger interaction between the oxygen-containing groups and oxidative fragments present on the GOF flake and the atoms at the perovskite surface. 
We quantified this trend using the flake--slab distance ($d_\mathrm{P-F}$) and the binding energy per atom ($E_{\mathrm{bind}/N}$) (Table~\ref{tab:TS2}).
GOF is positioned closer to the $\mathrm{CsPbI_3}$ surface than GF, and the $\mathrm{CsPbI_3}$/GOF heterostructure has the most negative $E_{\mathrm{bind}/N}$ among the studied systems. Since $E_{\mathrm{bind}/N}$ measures the overall stability of the optimised system relative to isolated atoms, this indicates that the GOF-containing heterostructure is the most strongly bound system considered here. However, this stronger binding is accompanied by the largest structural distortion. GF, in contrast, stabilises the slab while producing much smaller displacements of the perovskite atoms.
The average bond lengths analysis in  $\mathrm{CsPbI_3}$/GF heterostructure indicates that GF largely preserves the Pb--I framework of the $\mathrm{CsPbI_3}$ slab (Table~\ref{tab:perovskite_bond_lengths}). The Cs--I and Cs--Pb distances show only a weak shortening trend within the uncertainty of the data, suggesting local surface-coordination adjustment rather than substantial slab distortion.

We then extended the GF analysis to $\mathrm{MAPbI_3}$ and $\mathrm{CsSnBr_3}$ slabs to test whether this behaviour is specific to $\mathrm{CsPbI_3}$ (Figure~\ref{fig:s10}a,b; Table~\ref{tab:TS2}). In all cases, addition of GF makes $E_{\mathrm{bind}/N}$ more negative relative to the corresponding pristine slab, indicating stabilisation of the flake-containing systems. The $\mathrm{MAPbI_3}$/GF system shows the largest maximum displacement among the GF-containing slabs. However, the average Pb--I distance in $\mathrm{MAPbI_3}$/GF (3.123$\pm$0.043~\AA) remains comparable to that in $\mathrm{CsPbI_3}$/GF (3.139$\pm$0.061~\AA), indicating that this larger displacement is not associated with stronger distortion of the Pb--I framework. It likely reflects the additional orientational freedom of the MA cations and their local adjustment within the iodide cage. In contrast, the two inorganic-cation systems, $\mathrm{CsPbI_3}$/GF and $\mathrm{CsSnBr_3}$/GF, show more comparable and less pronounced surface distortions. 
Overall, GF provides a comparatively non-disruptive contact across the halide perovskite surfaces considered, stabilising the exposed surface while inducing smaller structural distortions than GOF.
This behaviour is relevant to graphene-containing active layers in polycrystalline perovskite films, where flakes may contact grain surfaces, grain boundaries, and weakly coordinated regions within the perovskite layer. The ability of GF to stabilise such regions without substantially distorting the inorganic framework provides a possible microscopic explanation for previous reports showing that pristine graphene or graphene nanostructures can improve perovskite crystallisation and film stability, including by suppressing iodide loss.~\cite{Li2018,Redondo2021,Zhao2018}


We next analysed how flake adsorption modifies the electronic structure of the perovskite slabs. Compared to the bulk system (Figure~\ref{fig:dft} d, middle-panel inset), the slab (Figure~\ref{fig:dft}d, middle panel) already exhibits changes in the density of states due to the presence of surface states and a shift of the Fermi level (Table~\ref{tab:TS2}). Adsorption of GF introduces additional states near the valence and conduction band edges and within the band-gap region, mainly originating from edge C atoms of the flake, as confirmed by the projected density of states (PDOS) analysis (Figure~\ref{fig:dft}d; Figure~\ref{fig:s11}). These states are weakly to moderately localised, as indicated by the electronic inverse participation ratio (IPR), where larger values correspond to stronger spatial localisation of the electronic states (Figure~\ref{fig:dft}d, left-panel inset; Figure~\ref{fig:s11}a,b, right-panel insets).
GOF gives a markedly different electronic response. In this case, flake-derived states form a denser manifold across much of the band-gap region and show substantially larger IPR values (Figure~\ref{fig:dft}d, right-panel inset), indicating stronger localisation. This behaviour is consistent with contributions from flake O atoms and edge C atoms in GOF, indicating that oxygen functionalisation makes the flake more electronically active. The spin-resolved electronic structure provides a further distinction: the GF-containing systems remain spin-unpolarised, whereas the $\mathrm{CsPbI_3}$/GOF system converges to a spin-polarised state with spin-channel asymmetry mainly in the band-gap region (Figure~\ref{fig:s8}). Thus, GF preserves a comparatively benign interface, while GOF is more likely to introduce trap-relevant electronic states.

The charge-density difference maps reveal the origin of this contrast (Figure~\ref{fig:dft}e; Figure~\ref{fig:s12}a). For GF-containing systems, the charge rearrangement remains mainly associated with the flake edges and the outer perovskite surface, indicating a surface-confined and weakly polar contact. This behaviour is observed not only for $\mathrm{CsPbI_3}$/GF, but also for the $\mathrm{MAPbI_3}$/GF and $\mathrm{CsSnBr_3}$/GF slabs (Figure~\ref{fig:s12}; Table~\ref{tab:TS2}). Although the magnitude of the vacuum-level shift and interface dipole depends on the perovskite composition, GF consistently modifies the surface energetics without producing the strongly localised, oxygen-derived electronic response observed for GOF.
GOF gives a qualitatively different response. The oxygen-containing groups and oxidative fragments bring the flake closer to the $\mathrm{CsPbI_3}$ surface and generate a stronger, oppositely oriented interface dipole (Table~\ref{tab:TS2}). Thus, GOF is not simply a more strongly binding analogue of GF, but creates a more polar and electronically active interface. 

Layer-resolved DOS projections onto Cs and I atoms of the surface and central CsI layers further show how far these changes penetrate through the $\mathrm{CsPbI_3}$ perovskite slab (Figure~\ref{fig:s9}a,b). The perturbation induced by GF within the perovskite is largely surface-confined, with only weak Cs- and I-derived gap-region peaks appearing in the surface layer and becoming negligible in the central layer. Additional spin-polarised density functional tight-binding (DFTB) calculations on nine-layer-thick $\mathrm{CsPbI_3}$ slabs interacting with two GFs confirm this, showing no gap states in the fourth CsI layer from the GF-contacting surface (Figure~\ref{fig:s15}). In contrast, interaction with GOF produces much stronger Cs- and I-derived gap states at the surface, which remain visible, although reduced, in the central layer. Thus, the GOF-induced electronic perturbation extends further through the slab thickness than that induced by GF.  

The macroscopically averaged electrostatic potential profiles provide the corresponding layer-resolved picture of the surface potential landscape (Figure~\ref{fig:dft}f,g; Figure~\ref{fig:s12}b,c). For $\mathrm{CsPbI_3}$/GF, the difference between the total electrostatic potential of the pristine slab and that of the slab interacting with GF shows pronounced local modulations near the flake-contacted surface layers. This is consistent with the metallic character of GF, which enables local screening and charge redistribution at the perovskite surface. These local changes are, however, distinct from the net electrostatic offset: despite the larger local modulation, $\mathrm{CsPbI_3}$/GF gives only a weak vacuum-level shift and a small interface dipole (Table~\ref{tab:TS2}). In the GOF case, the red-framed regions indicate a broader potential perturbation at the perovskite surface induced by the GOF, which is surrounded by stabilising oxidative debris. The green-framed outer CsI regions exhibit opposite electrostatic responses for GF and GOF, consistent with the opposite signs of the corresponding interface dipoles. The analysis of $\mathrm{MAPbI_3}$/GF and $\mathrm{CsSnBr_3}$/GF systems shows that this GF-induced electrostatic response is retained across different perovskite compositions, although its magnitude depends on the specific surface chemistry and lattice composition (Figure~\ref{fig:s12}b,c; Table~\ref{tab:TS2}). Among the systems considered, $\mathrm{CsPbI_3}$/GF shows the weakest net electrostatic perturbation, whereas $\mathrm{MAPbI_3}$/GF and $\mathrm{CsSnBr_3}$/GF exhibit larger but still surface-localised shifts. This supports the broader compatibility of GF with halide perovskite surfaces, while showing that the strength of the interfacial electrostatic tuning remains composition-dependent.

The work-function shifts place these electrostatic changes in an optoelectronic context. GF increases the work function of the perovskite slabs while maintaining a comparatively benign electronic response, whereas the GOF-induced work-function change is accompanied by stronger interfacial polarisation and localised gap-region states (Table~\ref{tab:TS2}). 
Such shifts are relevant for tuning local energy-level alignment at perovskite interfaces.~\cite{Canil2021} However, when work-function modification is accompanied by localised gap states and strong surface dipoles, it may favour trap-assisted losses, particularly when these interfaces are present within or close to the photoactive perovskite region.~\cite{Sherkar2017} From this perspective, GF provides a more favourable balance: it tunes the surface energetics while preserving a less disruptive electronic interface.

Taken together, our calculations demonstrate that the beneficial interaction with perovskites arises from the largely pristine character of GF: it stabilises exposed perovskite regions while causing only limited structural and electronic perturbation. By contrast, GOF binds more strongly but introduces larger distortions and localised gap states. 


\begin{figure}[!tb]
    \centering
    \includegraphics[width=0.99\columnwidth]{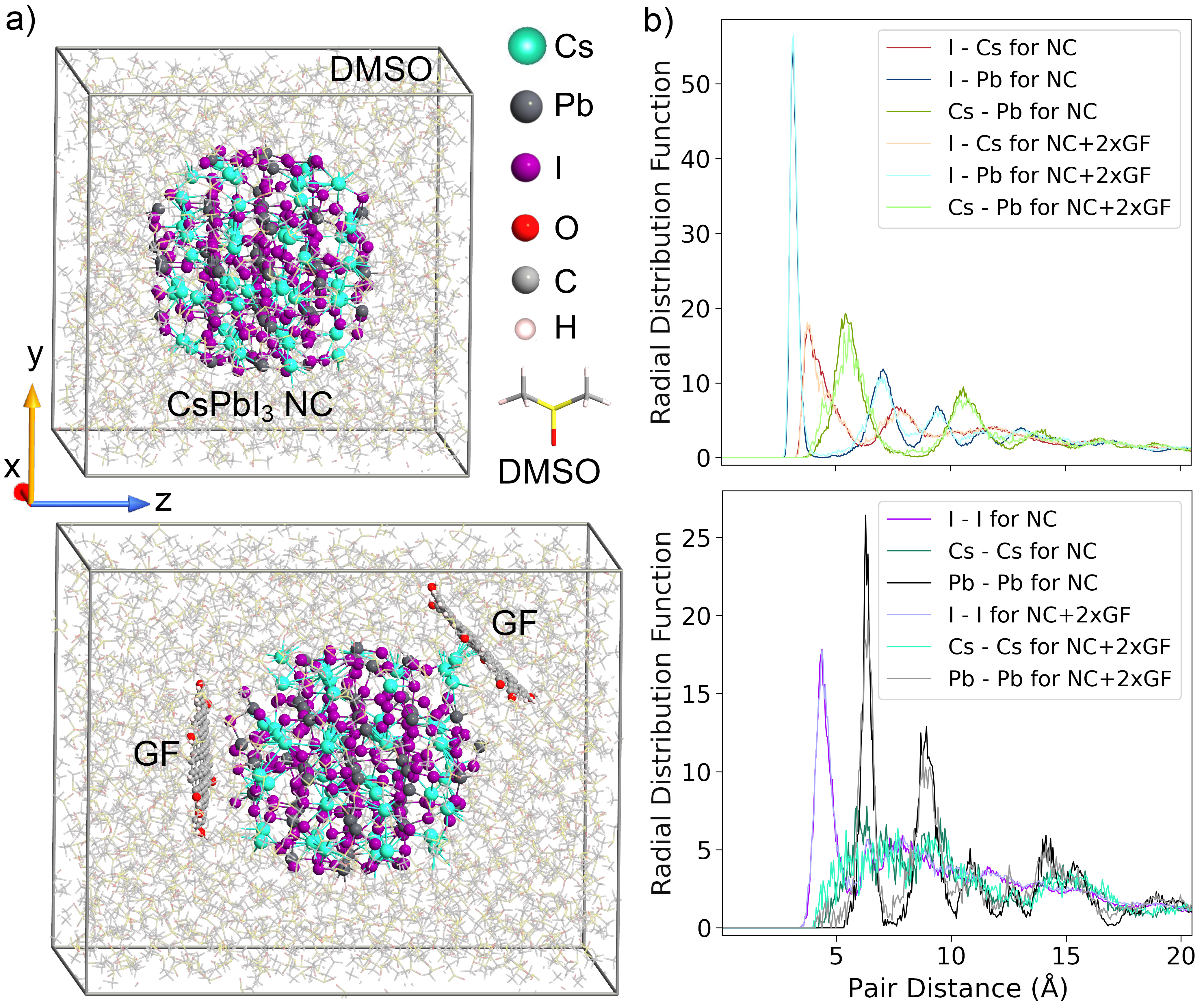}
    \caption{Computational evidence for GF-induced local reorganisation of $\mathrm{CsPbI_3}$ nanocrystals in DMSO.
    a) Final configurations from ML-MD/TFMC simulations of $\mathrm{CsPbI_3}$ nanocrystals (NCs) in small DMSO-filled simulation boxes, without GF (top, NC$^\mathrm{sbox}$) and with two GF (bottom, NC$_\mathrm{GF}^\mathrm{sbox}$). The two GF were initially placed above different NC facets: a PbI$_2$-terminated (100) facet and a Cs/I-rich (101) facet. b) Radial distribution functions (RDFs) calculated from the final 4~ps of the small-box DMSO simulations, comparing the NC and NC+2GF models. The upper panel shows I--Cs, I--Pb, and Cs--Pb pair correlations, while the lower panel shows I--I, Cs--Cs, and Pb--Pb pair correlations.
    }
    \label{fig:NC}
\end{figure}

To determine whether the GF--perovskite interactions identified for the slabs persist in a solvated nanoscale environment, we performed additional machine-learning molecular dynamics and time-stamped force-bias Monte Carlo (ML-MD/TFMC) simulations of a $\mathrm{CsPbI_3}$ nanocrystal (NC) interacting with two GFs in $N,N$-dimethylformamide (DMF) and dimethyl sulfoxide (DMSO) (Section~\ref{sec:MDMC} in the Supporting Information, Figure~\ref{fig:s14}, and Table~\ref{tab:perovskite_bond_lengths}). These solvents were selected because they are widely used as coordinating media in metal-halide perovskite precursor processing, including the preparation of $\mathrm{CsPbI_3}$ films.~\cite{BautistaQuijano2023, Wang2018} To quantify the GF--NC interactions, we calculated radial distribution functions (RDFs), which describe how frequently selected atom pairs occur at a given distance, with the first peak indicating the shortest preferred contact distance. The RDF analysis shows that the solvent is not a passive medium but actively modulates the GF--NC interactions (Figure~\ref{fig:s10}a). In these models, the two GFs were initially placed above different NC surface terminations, namely a PbI$_2$-terminated (100) facet and a Cs/I-rich (101) facet. In DMSO, they remain at comparable distances from the NC, whereas in DMF, the flake above the Cs/I-rich (101) facet moves farther from the perovskite NC. This behaviour is consistent with the different solvation environments generated by DMSO and DMF. Their donor numbers, which measure their ability to donate electron density to and coordinate Pb$^{2+}$-containing lead-halide species, differ~\cite{Petrov2021, Hamill2018} and may therefore change the interactions between GF and perovskite NCs. Further RDF analysis identified NC--GF contact motifs in DMSO. Oxygen-containing groups at the GF edges approach surface Pb atoms most closely, whereas basal-plane C atoms show the shortest-distance correlations with surface I atoms (cf. the positions of the first RDF peaks in Figure~\ref{fig:s10}b).  Comparison of DMSO simulations with and without GF shows that GF modifies both NC bond lengths and local pair correlations. In particular, in the presence of GF, Cs--Pb and Pb--Pb correlations shift slightly toward shorter distances, while the Pb--I correlations remain comparatively less affected (Figure~\ref{fig:NC}), consistent with the Cs--Pb and Pb--I bond-length analysis of the final NC configurations (Table~\ref{tab:perovskite_bond_lengths}). These short ML-MD/TFMC simulations of preformed perovskite NC therefore show that GF can interact simultaneously with Pb cations and I anions at the perovskite surface and reorganise the local NC environment without strongly disrupting the Pb--I framework.

\begin{figure}[!tb]
    \centering
    \includegraphics[width=0.98\columnwidth]{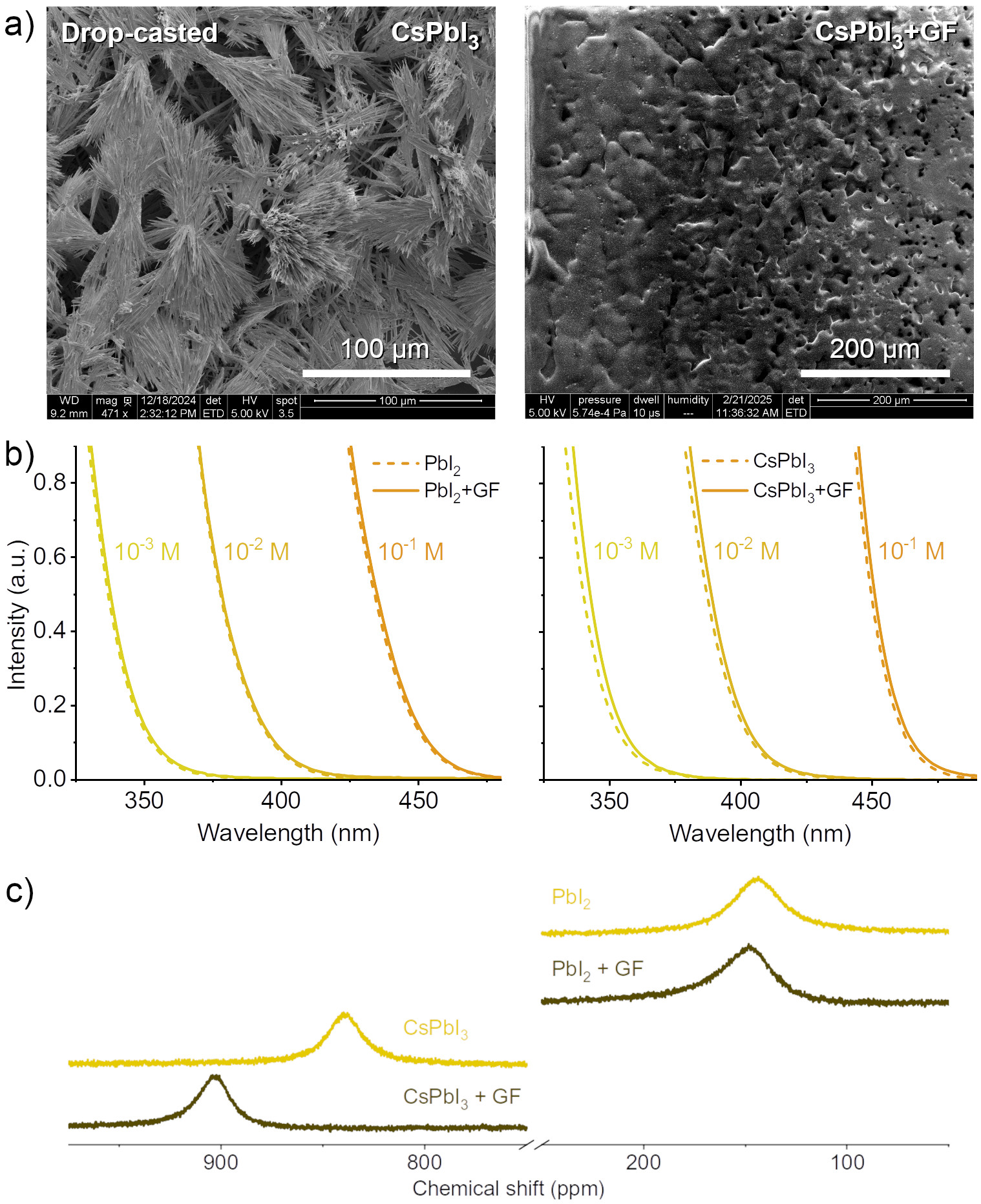}
    \caption{Experimental evidence for GF-induced precursor pre-organization and modified CsPbI$_3$ crystallization. a) SEM of drop-casted CsPbI$_3$ films from 0.1~M solutions with and without GF. b) UV-vis spectra of PbI$_2$ and CsPbI$_3$ solutions at different concentrations with and without GF. c) $^{207}$Pb NMR spectra of PbI$_2$ and CsPbI$_3$ with and without GF.}
    \label{fig:FigE1}
\end{figure}

Therefore, we set out to test experimentally whether pristine GF can influence the organisation of Pb/I-containing precursor species in solution and the subsequent crystallisation of metal halide perovskites. We anticipated that GF could promote precursor organisation without the substantial structural perturbations predicted for GOF.
To test this prediction, we prepared Pb-based perovskite precursor solutions with and without GF. We chose CsPbI$_3$ as a model composition because of its chemical simplicity and direct correspondence to the calculated systems.
We first prepared GF dispersions in DMSO, a solvent capable of dissolving and processing perovskite films, for which our simulations predicted closer GF--perovskite NC contacts than in DMF, and chloroform (CHCl$_3$), a solvent commonly used for stable GF dispersions.~\cite{ONeill2011}
Both solvents effectively dispersed GF at 1 wt\% (Figure \ref{fig:FigSE1}), after which the dispersions were filtered. The resulting clear solutions were then used to dissolve the perovskite precursors, CsI and PbI$_2$. While DMSO readily dissolves the precursors, CHCl$_3$ does not.
Accordingly, we used the DMSO-based dispersions for subsequent solution studies and prepared drop-cast films to examine the influence of GF on crystallisation. A 0.1~M perovskite solution was used to maximise the relative GF content and amplify its effect on the perovskite precursors. Scanning electron microscopy (SEM) revealed a clear effect of GF on perovskite crystallisation (Figure~\ref{fig:FigE1}a). Films prepared without GF contained disordered, needle-like CsPbI$_3$ crystals, whereas those prepared from the filtered GF-containing dispersions exhibited compact, grain-based microstructures resembling conventional polycrystalline perovskite thin films.
This change suggests a significant interaction between GF and the perovskite system and indicates a potential positive role of GF in promoting a more organised perovskite crystallisation.

To investigate these interactions, we characterised the solutions by UV-vis spectroscopy at concentrations of 10$^{-3}$, 10$^{-2}$ and 10$^{-1}$ (Figure~\ref{fig:FigE1}b). No notable effect of GF is observed for PbI$_{2}$ dispersions; however, GF-containing CsPbI$_3$ dispersions consistently exhibit a redshift of 1--2\,nm. We excluded any relevant contribution from GF as too far into the blue (Figure~\ref{fig:FigSE2}). The observed redshift indicates enhanced Pb$^{2+}$ coordination by iodide, reflecting a higher degree of pre-organisation of the perovskite precursor materials in solution. This effect is less pronounced at higher perovskite concentrations, suggesting it is more relevant at lower levels of precursor organisation. We obtained further evidence by $^{207}$Pb NMR characterisation of the dispersions, which offers relevant information on the environment of the metal cation Pb$^{2+}$ (Figure~\ref{fig:FigE1}c).~\cite{Hu2025} The chemical shift of PbI$_2$ remains comparable; however, the CsPbI$_3$ signal shifts 65\,ppm downfield after GF addition, which indicates a lower electron density around the metal. This shift implies decreased solvation of Pb$^{2+}$ by DMSO, the most electron-donating species in the system, and correspondingly a higher degree of Pb-I coordination. Thus, these results indicate that GF promotes precursor pre-arrangement and enhances structural organisation in solution. 


Given the potential of few-layer pristine GFs to enhance perovskite film formation, we tested the potential of perovskite-GF mixtures to yield higher-quality thin films and devices. We built solar cells based on ITO/BCZ/perovskite/C$_{60}$/BCP/Ag architecture, with a perovskite composition of Cs$_{0.05}$(FA$_{0.95}$MA$_{0.05}$)$_{0.95}$Pb(I$_{0.95}$Br$_{0.05}$)$_{3}$, finding slight improvements when in the presence of any concentration of GF, but a relatively comparable performance (Figure~\ref{fig:FigSE3}). The improvement was mainly due to impressive FF values over 0.82; however, voltages decreased slightly. For the different concentrations, different percentages (10\%, 50\%, 100\%) of GF-saturated DMSO and pure DMSO were used to dissolve the perovskite precursors.

After confirming that GF could be incorporated into a conventional Pb-only perovskite device without compromising its photovoltaic performance, we examined whether it could be particularly beneficial for more challenging perovskite compositions that typically yield lower-quality films and are more prone to fabrication-related defects.
Among the perovskite types that suffer from a challenging crystallisation, we chose mixed Sn--Pb perovskites due to their narrow bandgap and potential for single junction and tandem solar cells. Due to their wide particle size distribution in solution,~\cite{Pascual2026} these perovskites exhibit unordered nucleation, a challenge that GF could help mitigate. We investigated the potential effect of GF in the thin film properties of Cs$_{0.1}$FA$_{0.6}$MA$_{0.3}$Sn$_{0.5}$Pb$_{0.5}$I$_{3}$ perovskite. We incorporated GF to the perovskite precursor solution based on the Sn--Pb composition and prepared thin films by spin-coating (Figure~\ref{fig:FigSE4}). The characterisation of the films by SEM shows top-view images displaying compact multigrain structures, with the presence of compounds of a different nature or unreacted material on top of the control films (Figure~\ref{fig:FigSE5}). In contrast, the incorporation of GF suppresses these residues, with the effect becoming more pronounced at higher concentrations. Grain-size distribution analysis of top-view images reveals an overall increase in average grain size upon GF incorporation, a beneficial effect that reduces grain boundaries and associated surface defects. (Figure~\ref{fig:FigSE6}). Structural characterisation of the films by XRD analysis shows the main representative perovskite signals (Figure~\ref{fig:FigSE7}). It also shows the presence of the peak at $13^\circ$, attributable to the presence of PbI$_2$, therefore confirming the presence of unreacted material in control films. With higher GF concentrations, this peak disappears, confirming the ability of GF to promote an efficient mixing within the dispersions and influence positively the crystallisation of the perovskite.

\begin{figure}[!th]
    \centering
    \includegraphics[width=0.48\textwidth]{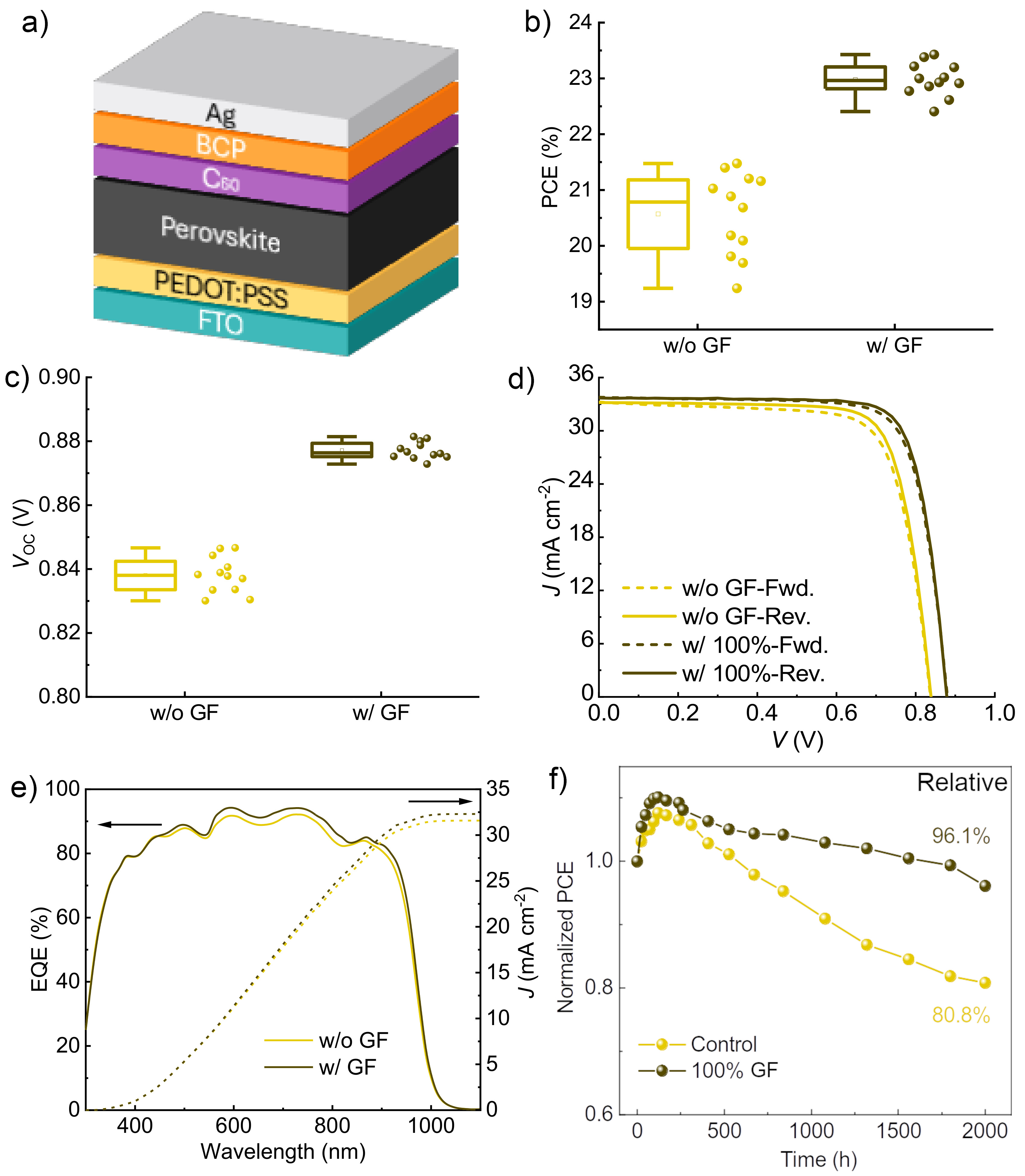}
    \caption{GF-enhanced photovoltaic performance of mixed Sn--Pb perovskite solar cells a) Device architecture used for Sn--Pb PSCs. b) Distribution of PCE and c) $V_\text{OC}$ values. d) J--V curves for the champions' devices with and without the optimum concentration of GF. e) EQE spectra for Sn--Pb devices with different concentrations of GF. f) Evolution of the power conversion efficiency (PCE) of control and 100\% GF-containing perovskite solar cells during ageing. }
    \label{fig:device}
\end{figure}

We evaluated the PV performance of the improved Sn--Pb perovskite films in devices. We incorporated the films into the common p-i-n architecture for these mixed metal compositions:  FTO/PEDOT:PSS/perovskite/C$_{60}$/BCP/Ag (Figure~\ref{fig:device}). The control devices show a power conversion efficiency (PCE) ranging from 19 to 21\% (Figure~\ref{fig:device}b and ~\ref{fig:FigSE8}). The best efficiency obtained was 21.5\%, with a short-circuit current density (\textit{J}$_\mathrm{SC}$) of 33.2~mA~cm$^{-2}$, an open-circuit voltage (\textit{V}$_\mathrm{OC}$) of 0.84~V, and a fill factor (FF) of 0.77. In contrast, the GF-containing devices exhibited improved values ranging from 22 to 23\%, with much narrower distribution. The record GF-containing device presented PCE of 23.7\%, with \textit{J}$_\mathrm{SC}$ of 33.4~mA~cm$^{-2}$, \textit{V}$_\mathrm{OC}$ of 0.88~V, and FF of 0.80. All PV parameters improve, resulting in enhanced PCE valuess, although the most significant improvements was in the Voc (Figure~\ref{fig:device}c). Figure~\ref{fig:device}d shows the \textit{J}--\textit{V} curves of the champion control and target devices, also displaying the reduced hysteresis for the GF-containing device (0.037 vs 0.021). The external quantum efficiency (EQE) measurements showed integrated \textit{J}$_\mathrm{SC}$ values of 31.6 and 32.3~mA~cm$^{-2}$ for control and target devices, respectively, confirming the validity of the values obtained from the \textit{J}--\textit{V} curves (Figure~\ref{fig:device}e). The stabilised power output (SPO) showed efficiency values of 21.8 and 23.7\% for control and target devices, respectively (Figure~\ref{fig:FigSE10}). To evaluate the potential influence of GF inside perovskite thin films on device stability, we tracked the evolution of solar cells stored in N$_2$-filled glove boxes at 25~$^o$C (Figure~\ref{fig:device}f, Figure~\ref{fig:FigSE9}). The control device maintained 80\% of its initial efficiency after 2000 h. Meanwhile, the GF-containing device experienced only a slight drop in efficiency, maintaining 96\% of its initial efficiency over the same period. This result demonstrates the enhanced robustness provided by GF to perovskite solar cells, and thus the ability of true 2D amphiphilic materials to stabilise metal halide perovskite surfaces.

\section{Conclusions}
In summary, we establish that pristine graphene flakes (GF) and graphene oxide flakes (GOF) interact with halide perovskites in fundamentally different ways. DFT calculations show that GOF binds more strongly but induces larger structural distortions, stronger interfacial polarisation, and localised gap states. In contrast, pristine GF forms stable contacts that largely preserve the perovskite framework and confine the electronic perturbation to the surface across $\mathrm{CsPbI_3}$, $\mathrm{MAPbI_3}$, and $\mathrm{CsSnBr_3}$. ML-MD/TFMC simulations further identify complementary contacts between GF edge O atoms and surface Pb atoms and between basal-plane C atoms and surface I atoms, while showing that the solvent modulates these interactions without strongly disrupting the Pb--I framework. Consistent with this atomistic picture, UV--vis and $^{207}$Pb NMR measurements indicate that GF reorganises the Pb/I-containing precursor environment, promoting more organised crystallisation and improved film morphology. Although only modest effects are observed in Pb-only devices, the benefit becomes pronounced for mixed Sn--Pb perovskites, where GF suppresses residual phases, increases grain size, raises the PCE from 21.5\% to 23.7\%, and improves PCE retention after 2000~h from 80\% to 96\%. Together, these results provide a multiscale mechanism linking pristine GF chemistry to precursor organisation, crystallisation, and device performance, and establish GF as a promising additive for difficult-to-process halide perovskites.

\section*{Author contributions}
The concept of this work was developed jointly by K.Z.M. and J.P. K.Z.M. led and coordinated the theoretical part of the study, while J.P. led and coordinated the experimental part. K.Z.M and J.P. together defined the integration of the experimental and theoretical parts. M.B.T. performed the initial DFT optimisation of the heterostructure configurations under the supervision of K.Z.M. O.E.-R. performed the DFT and DFTB simulations under the supervision of K.Z.M. K.Z.M. performed the ML-MD/TFMC simulations. O.E.-R., K.Z.M., and D.Bx. analysed the simulation results. D.Ba. performed the preparation and characterization of precursor solutions and drop-casted films under the supervision of J.P. A.Ch. performed the microscopy characterisation of the film samples. S.H-T.C. performed NMR analyses. J.P. supervised the experimental work and collected and analysed the experimental data. Y.G. performed solar cell fabrication and characterization under the supervision of M.L. O.E.-R., D.Ba., J.P., and A.L.-R. prepared the first version of the manuscript, which was subsequently revised by K.Z.M., D.Bx., A.L.-R., and J.P. Funding for the study was secured by K.Z.M., J.P., and A.L.-R. All authors discussed the results and contributed to the final manuscript.

\section*{Conflicts of interest}
There are no conflicts to declare.

\section*{Data availability}
The data supporting this article are included in the Electronic Supplementary
Information (ESI\dag).

\section*{Acknowledgements}
The authors thank Andrey Chuvilin for performing the SEM measurements.
O.E.-R. and K.Z.M acknowledge the technical and human support provided by the DIPC Supercomputing Center, Spain and  by the Interdisciplinary Centre for Mathematical and Computational Modelling at the University of Warsaw, Poland (Grant No. GC84-32). 
K.Z.M also would like to thank the European Commission (Marie Sklodowska-Curie Cofund Programme; grant no. H2020-MSCA-COFUND-2020-101034228-WOLFRAM2) for funding this research.  J.P. acknowledges support from the European Commission (Marie Sklodowska-Curie Cofund Programme; grant no. H2020-MSCA-COFUND-2020-101034297-E4F).
K.Z.M. and O.E.-R. are grateful to the Agencia Estatal de Investigacion, Ministerio de Ciencia e Innovacion, Spain for supporting this research under Proyectos de Generacion de Conocimiento 2022 program, PID2022-139776NB-C65. K.Z.M also would like to aknowledge the support from the Ramón y Cajal grant RYC2024-051436-I, funded by MICIU/AEI/10.13039/501100011033 and by ESF+.
The work of D.Bx. and O.E.-R. is supported by the Grant PID2024-162933NB-I00 (QUILL) funded by MICIU/AEI/10.13039/501100011033 by ERDF/EU. D.Bx. acknowledge the financial support of the Basque Government’s Department of Education through project PIBA\_2023\_1\_0007 (STRAINER), the IKUR Strategy under the collaboration agreement between the Ikerbasque Foundation and DIPC, on behalf of the Basque Government’s Department of Education and the Gipuzkoa Provincial Council, within the QUAN-000021-01 project; and the ``Artificial Quantum Matter: From 2D Materials to Spin Lattice Systems" - BBVA Foundation Fundamentos Program 2024".
D. Ba., and A.L.-R. would also like to thank the National Science Centre, Poland (PRELUDIUM BIS 4 2022/47/O/ST5/01336) for funding.
S.H.T.C. gratefully acknowledges funding from the Ministry of Science and Innovation of Spain under Ayudas Ramón y Cajal (RYC2022-035578-I) and the project PID2023-151880OB-C32 funded by MICIU/AEI/10.13039/501100011033.

\FloatBarrier
\clearpage
\thispagestyle{plain} 

\begin{center}
	\LARGE \textbf{Supporting Information} \\[0.5em]
	
	\Large
	for \\[0.5em]
	
	\Large{\title{Understanding Graphene–Perovskite Interactions: From Flake Chemistry to Crystallisation and Solar Cell Performance}
		\color{black}{}}
\end{center}

\vspace{0cm}

\setcounter{section}{0}
\setcounter{subsection}{0}
\setcounter{figure}{0}
\setcounter{table}{0}
\renewcommand{\thesection}{S\arabic{section}}
\renewcommand{\thesubsection}{S\arabic{section}.\arabic{subsection}}
\renewcommand{\thefigure}{S\arabic{figure}}
\renewcommand{\thetable}{S\arabic{table}}

\section{Materials and Methods}

\subsection{Calculation Details}
\subsubsection{Density Functional Theory calculations}

To investigate the structural and electronic properties of bulk cubic $\mathrm{CsPbI_3}$, cubic $\mathrm{CsSnBr_3}$, orthorhombic $\mathrm{MAPbI_3}$, and their corresponding slab models, we performed spin-polarized density functional theory (DFT)~\cite{hohenberg1964, kohn1965} calculations within the generalized gradient approximation (GGA)~\cite{perdew1996}, using the PBEsol exchange--correlation functional~\cite{perdew2008} as implemented in the \textsc{SIESTA} package~\cite{soler2002, ordejon1996}. 
The slab models were constructed from (001)-type surfaces with AX-type terminations, namely CsI-terminated $\mathrm{CsPbI_3}$, CsBr-terminated $\mathrm{CsSnBr_3}$, and MAI-terminated $\mathrm{MAPbI_3}$ surfaces, consistent with previous surface-stability studies of halide perovskites\cite{Marronnier2018,Seidu2021,Haruyama2014,Chen2023,Ma2018}.
The PBEsol functional was chosen for its improved accuracy in reproducing equilibrium lattice parameters of crystalline solids compared to the standard PBE formulation, making it particularly suitable for halide perovskites. All calculations employed a double-$\zeta$ plus polarization (DZP) numerical atomic orbital basis set in combination with norm-conserving pseudopotentials, ensuring a reliable balance between computational efficiency and accuracy for large periodic systems.

Slab models of halide perovskites were computed under periodic boundary conditions, with a sufficiently large vacuum layer to eliminate spurious interactions between periodic images. 
For bulk calculations, the Brillouin zone was sampled using Monkhorst--Pack~\cite{monkhorst1976} $k$-point grid of 10$\times$ 10 $\times$ 10, whereas slab geometries  employed  a 4$\times$ 4 $\times$ 1 grid.
Band structure calculations for the cubic bulk $\mathrm{CsPbI_3}$ system were performed along the high-symmetry path $\Gamma$--X--M--$\Gamma$--R--X in the Brillouin zone.

The real-space integration grid was set with a mesh cutoff of 550\,Ry for bulk systems and 450\,Ry for slabs, chosen to ensure convergence of the total energies and forces. All atomic structures were relaxed until the maximum residual force was less than 0.001\,eV/\AA for the bulk systems and 0.02\,eV/\AA for the slabs, with a maximum stress variation below 0.001\,GPa. The self-consistent field iterations converged when the total energy change was less than 10$^{-5}$\,eV and the density matrix variation was below 10$^{-4}$ per iteration. 
All calculations were performed in vacuum.

The electronic inverse participation ratio (IPR) was calculated to quantify the degree of localization of the electronic eigenstates. 
Within the localized atomic orbital (LCAO) framework implemented in QuantumATK, the IPR for the $n$-th eigenstate is defined as

\begin{equation}
	p^{(n)} = \sum_{i=1}^{N} \left| q_i^{(n)} \right|^2 ,
\end{equation}

where

\begin{equation}
	q_i^{(n)} = \sum_{\mu \in i, \nu} \mathrm{Re}\left[S_{\mu\nu} c_{\mu n}^{*} c_{\nu n}\right].
\end{equation}

Here $N$ is the number of atoms in the system, $c_{\mu n}$ are the expansion coefficients of the $n$-th Kohn--Sham wavefunction in the LCAO basis, and $S_{\mu\nu}$ is the overlap matrix between basis orbitals. 
The IPR provides a measure of the spatial localization of electronic states: fully delocalized states scale as $1/N$, whereas larger IPR values indicate increasingly localized states, such as defect or impurity states \cite{Simdyankin2004}. 
The calculations were performed using the QuantumATK simulation platform \cite{Smidstrup2019}.

The perovskite slab and flake distance ($d_{P-F}$) was estimated from the atomic coordinates. Specifically, the vertical separation ($z$) between the surface atoms of the perovskite slab and the closest atoms of the flakes was measured. Because the flakes are not perfectly planar, the distance varies across the interface; therefore, a range of values is reported in Table~\ref{tab:TS2} corresponding to the minimum and maximum distance between the flake and the perovskite surface.

The total electrostatic potential ($V_{\mathrm{H}} + V_{\mathrm{ion}}$) was analysed using the \textsc{Macroave} code, which performs planar averaging along the surface-normal direction followed by macroscopic averaging of the resulting profile. In this approach, the microscopic electrostatic potential $V(\mathbf{r})$ obtained from the DFT calculations is first averaged over planes parallel to the surface according to
\begin{equation}
	\bar{V}(z)=\frac{1}{S}\int_S V(x,y,z)\,dx\,dy ,
\end{equation}
where $S$ is the surface area of the unit cell. The remaining oscillations associated with the atomic structure are then removed by applying a macroscopic averaging procedure based on convolution with step-function filters along the $z$ direction. This procedure yields a smooth electrostatic potential profile across the slab, allowing identification of the vacuum level and enabling the estimation of work functions for the studied systems. The method follows the macroscopic averaging formalism introduced by Baldereschi \textit{et al.} and later applied to interface electrostatics in first-principles calculations.\cite{Baldereschi1988,Colombo1991,JunqueraMacroave}

The interface dipole per unit area ($\upmu/S$) was estimated from the shift of the vacuum electrostatic potential induced by the adsorption of graphene-based flakes on the perovskite slabs. The vacuum level $V_{\mathrm{vac}}$ was extracted from the macroscopic electrostatic potential in the vacuum region.
For $\mathrm{CsPbI_3}$ and $\mathrm{CsSnBr_3}$, $V_{\mathrm{vac}}$ was taken from the flat plateau of the potential profile. For orthorhombic $\mathrm{MAPbI_3}$, where the vacuum potential did not form a perfectly flat plateau, $V_{\mathrm{vac}}$ was instead estimated as the average electrostatic
potential within the vacuum region. The vacuum-level shift $\Delta V_{\mathrm{vac}}$ was then defined relative to the corresponding pristine slab, and the interface dipole per unit area was calculated as
\begin{equation}
	\frac{\mu}{S} = \varepsilon_0 \Delta V_{\mathrm{vac}},
\end{equation} where $\varepsilon_0$ is the vacuum permittivity.

The work function ($\Phi$) was obtained from the difference between the vacuum electrostatic potential and the Fermi level according to
\begin{equation}
	\Phi = eV_{\mathrm{vac}} - E_F
\end{equation}
where $V_{\mathrm{vac}}$ is the vacuum level determined from the plateau of the macroscopic electrostatic potential in the vacuum region and $E_F$ is the Fermi energy obtained from the DFT calculation.

The binding energy per atom ($E_{\mathrm{bind}/N}$) was calculated from the total energies obtained from the DFT calculations according to
\begin{equation}
	E_{\mathrm{bind}/N} = \frac{E_{\mathrm{tot}} - \sum_i N_i E_i}{N_{\mathrm{tot}}}
\end{equation}
where $E_{\mathrm{tot}}$ is the total energy of the system, $N_i$ is the number of atoms of species $i$, $E_i$ is the reference energy of the isolated atom of species $i$, and $N_{\mathrm{tot}}$ is the total number of atoms in the system. 

\subsubsection{Additional DFTB Calculations}

Additional calculations were carried out for cubic bulk $\mathrm{CsPbI_3}$, the isolated graphene flake, and three-layer and nine-layer thick $\mathrm{CsPbI_3}$ slabs interacting with two graphene flakes using the spin-polarised density functional tight-binding (DFTB) method~\cite{elstner1998, frauenheim2002}. The PTBP parameter set~\cite{cui2024}, specifically developed to enhance the accuracy of band gaps and structural parameters in semiconductors and insulators, was employed. 
For cubic bulk $\mathrm{CsPbI_3}$, calculations utilised a real-space mesh cut-off of 225\,Ha, a maximum interaction range of 10\,\AA, and a Monkhorst–Pack $k$-point grid of 10 $\times$ 10 $\times$ 10. The slab models and the isolated graphene flake, a real-space mesh cut-off of 15\,Ha, a maximum interaction range of 12\,\AA\ were used.  The slabs employed a Monkhorst–Pack $k$-point grid of 4 $\times$ 4 $\times$ 1, whereas the isolated graphene flake was treated as a molecular system with $\Gamma$-point sampling only.
Structural optimisation was deemed converged when the maximum residual force was below 0.01\,eV/\AA.
The DFTB formalism, derived from DFT within the PBE approximation, expands the total energy to second order in charge-density fluctuations, retaining much of the accuracy of full DFT at significantly lower computational cost. The PTBP parameterisation includes element-specific corrections and an improved treatment of exchange–correlation effects compared with earlier Slater–Koster sets. 
DFTB/PTBP was employed for the full geometry optimisation of selected systems. Its use was necessary for the nine-layer slab due to the prohibitive cost of full DFT, while calculations for cubic bulk and the three-layer structure were performed for benchmarking against DFT results. For electronic structure calculations, the $k$-point sampling of slabs was increased to 8 $\times$ 8 $\times$ 1.

\subsubsection{Benchmark calculations for cubic bulk $\mathrm{CsPbI_3}$}

\begin{figure}[h!tb] 
	\centering
	\includegraphics[width=0.9\textwidth]{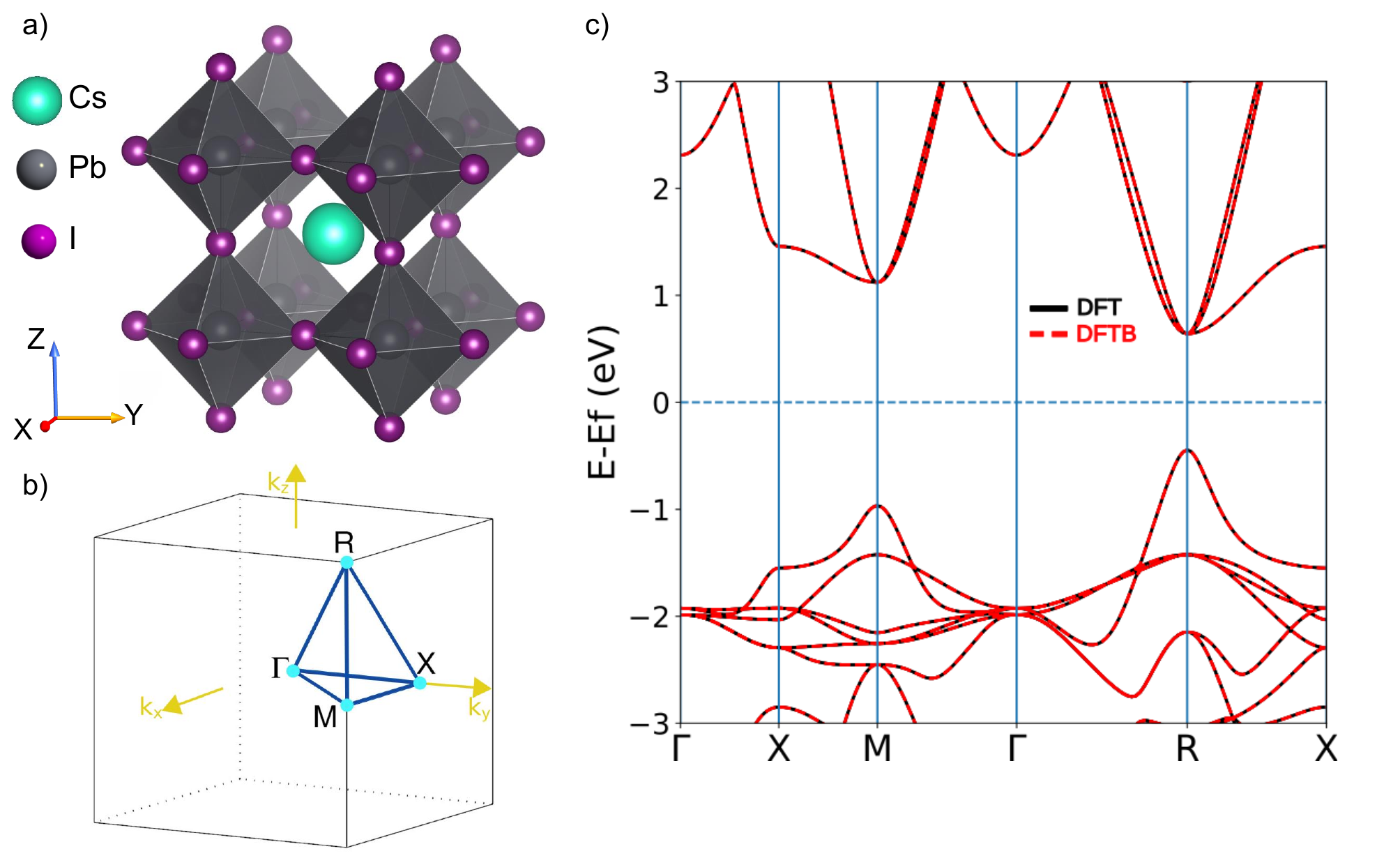} 
	\caption{a) Schematic lattice structure of bulk $\mathrm{CsPbI_3}$ in the cubic phase, 
		b) first Brillouin zone of the cubic system [$\Gamma$--X--M--$\Gamma$--R--X],  c) bandstructure of bulk $\mathrm{CsPbI_3}$ calculated at the DFT/PBEsol (black) and DFTB/PTBP (red) levels of theory.
	}
	\label{fig:s1}
\end{figure}

As shown in Figure~\ref{fig:s1}, the comparison of the band structure reveals that DFT/PBEsol and DFTB/PTBP predict qualitatively similar electronic dispersion near the band edges, and both methods confirm a direct band gap at the R point. Quantitatively, the DFTB/PTBP gap is marginally larger than the DFT/PBEsol value, consistent with the PTBP parameter set's optimization towards improved band-gap accuracy for semiconductors and insulators. In the conduction and valence band regions farther from the Fermi level, DFTB/PTBP reproduces the main dispersion features of DFT/PBEsol but exhibits small deviations in curvature, which may lead to small differences in the effective masses.

\begin{table}[h!tb]
	\centering
	\caption{Lattice parameters of bulk cubic $\mathrm{CsPbI_3}$ obtained from DFT/PBEsol, DFTB/PTBP, together with the experimental values reported in Ref.~\cite{TROTS20082520}.}
	\begin{tabular}{cccc}
		\hline
		\textbf{Lattice parameters} & \textbf{DFT (PBEsol)} & \textbf{DFTB (PTBP)}  & \textbf{Experimental ~\cite{TROTS20082520} }\\
		\hline
		$a$ [\AA]        & 6.216 & 6.332 & 6.289 \\
		$b$ [\AA]        & 6.216 & 6.332 & 6.289 \\
		$c$ [\AA]        & 6.216 & 6.332 & 6.289 \\
		$\alpha$ ($^{\circ}$) & 90    & 90 & 90  \\
		$\beta$ ($^{\circ}$)  & 90    & 90  &  90\\
		$\gamma$ ($^{\circ}$) & 90    & 90  &  90\\
		\hline
	\end{tabular}
	\label{tab:TS1}
\end{table}

The optimized lattice parameters (Table~\ref{tab:TS1}) of cubic $\mathrm{CsPbI_3}$ from DFT/PBEsol and DFTB/PTBP show close agreement. DFT predicts $a = b = c = 6.216$\,\AA, while DFTB produces slightly larger values of $a = b = c = 6.332$\,\AA, corresponding to a uniform overestimation of approximately 1.87\,\%. This minor discrepancy reflects the common tendency of certain tight-binding parameterizations to slightly expand the equilibrium volume. Both approaches preserve cubic symmetry, with calculated $\alpha$, $\beta$, and $\gamma$ angles fixed at 90$^{\circ}$. The calculated band gaps are slightly underestimated compared to experimental values, consistent with the known limitations of these methods. In general, DFTB/PTBP accurately reproduces the structural and electronic properties of cubic $\mathrm{CsPbI_3}$ in comparison with the corresponding DFT/PBEsol results.

\subsubsection{Benchmark calculations for isolated graphene flake}

\begin{figure}[h!tb] 
	\centering
	\includegraphics[width=0.9\textwidth]{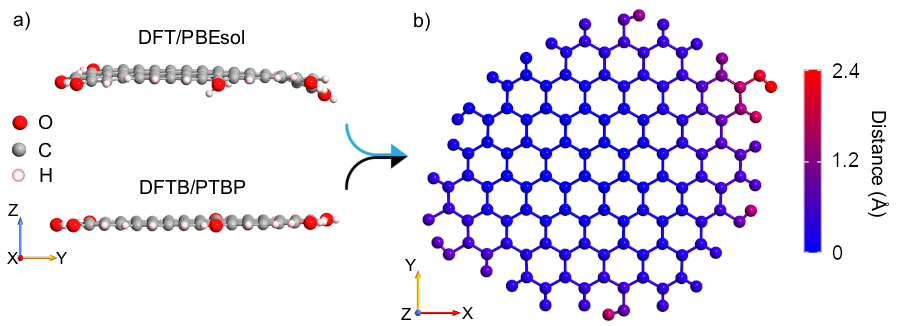} 
	\caption{a) Side view of the fully optimised isolated graphene flake obtained at the DFT/PBEsol and DFTB/PTBP levels of theory. b) Top view of atomic displacement map showing structural differences between the fully optimised DFTB/PTBP and DFT/PBEsol graphene flakes (DFT/PBEsol taken as the reference). The DFTB/PTBP structure is presented in a ball-and-stick representation, with atoms coloured according to the absolute displacement relative to the DFT geometry, ranging from blue (no displacement) through purple to red (maximum displacement).}
	\label{fig:s2}
\end{figure}

As shown in Figure~\ref{fig:s2}, the DFT/PBEsol optimisation yields a slightly curved, non-flat configuration of the graphene flake, whereas the DFTB/PTBP approach results in a more planar and overall flatter geometry. The structural differences between the two configurations are predominantly localised at the flake edges, particularly around the oxygen-containing functional groups. The average atomic displacement between the DFT and DFTB geometries is 0.40332 \AA, with a maximum deviation of 2.4392 \AA.

\subsubsection{Benchmark calculations for  $\mathrm{CsPbI_3}$ (5$\times$5$\times$3) slab/graphene flake heterostructure}

\begin{figure}[h!tb] 
	\centering
	\includegraphics[width=0.85\textwidth]{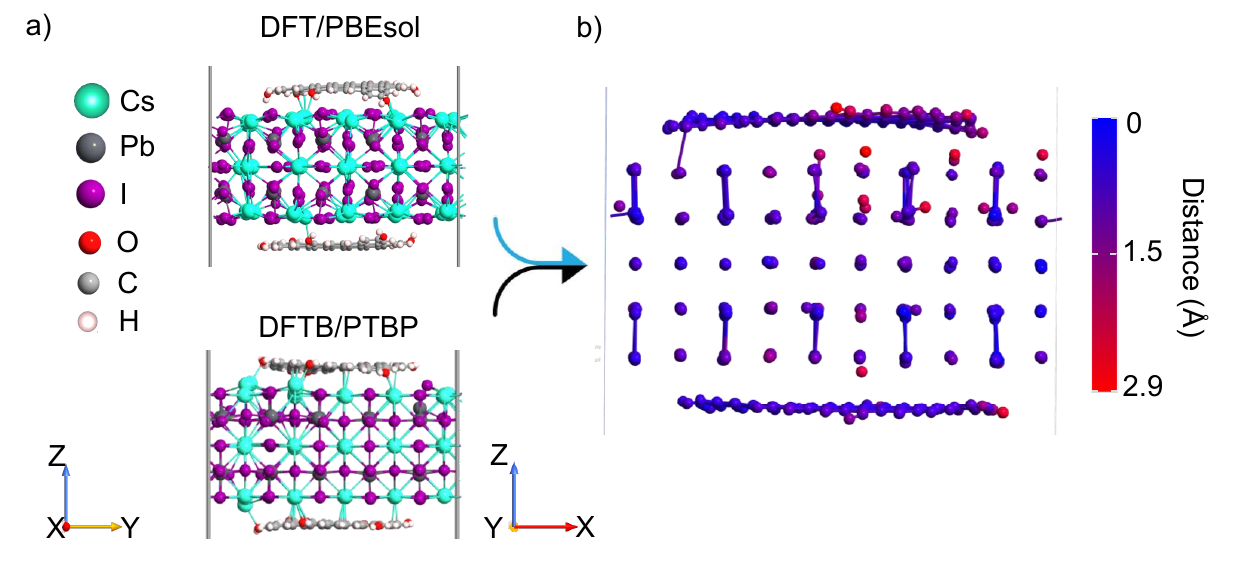} 
	\centering
	\includegraphics[width=0.80\textwidth]{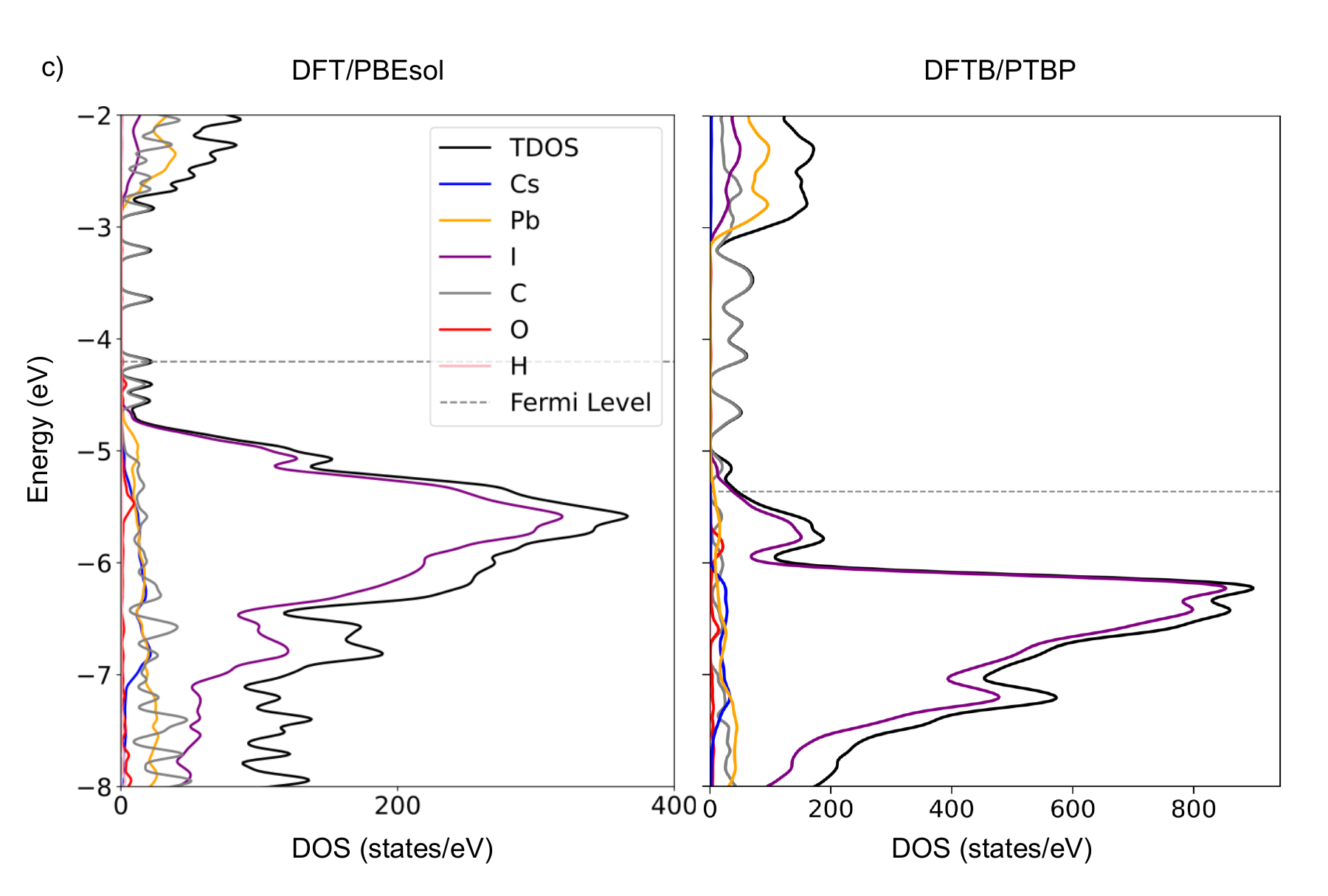} 
	\caption{
		a) Side view of the fully optimized $\mathrm{CsPbI_3}$ $(5 \times 5 \times 3)$/GF heterostructure obtained at the DFT/PBEsol and DFTB/PTBP levels of theory. 
		b) Atomic displacement map highlighting the structural differences between the fully optimized DFTB/PTBP and DFT/PBEsol geometries (DFT/PBEsol taken as the reference). The DFTB/PTBP structure is displayed in a ball representation, with atoms colored according to the absolute displacement relative to the DFT geometry, ranging from blue (no displacement) to red (maximum displacement). Interatomic connections are shown only for distances below the visualization cutoff (d$_{\mathrm{ij}}$) used for bond detection ($d_{ij}=1.1(R_i+R_j))$, where $R_i$ and $R_j$ are the covalent radii of the atoms). The absence of a connecting line therefore reflects distances exceeding this threshold rather than the absence of bonding interactions.
		c) Density of states of the $\mathrm{CsPbI_3}$ $(5 \times 5 \times 3)$ slab/GF heterostructure calculated at the DFT/PBEsol and DFTB/PTBP levels of theory, showing the total density of states and contributions projected onto Cs, Pb, I, all C, O, and H atoms. The dashed horizontal line indicates the Fermi level.
	}
	\label{fig:FigSTbG}
\end{figure}

The optimized structures (Figure~\ref{fig:FigSTbG}a) show that both approaches reproduce a similar geometry of the slab–flake interface, with a slab thickness of $\sim$1.2 nm at the DFT/PBEsol level and $\sim$1.28 nm for the DFTB/PTBP structure, indicating a comparable description of the slab thickness. The graphene flake remains adsorbed above the perovskite surface with an interlayer distance between the flake and the slab ranging from 2.65–3.33~\AA\ at the DFT/PBEsol level and 3.06–3.5~\AA\ at the DFTB/PTBP level. The displacement map (Figure~\ref{fig:FigSTbG}b), calculated relative to the DFT/PBEsol structure, indicates that the structural differences are moderate and mainly localized in the lower part of the perovskite slab and at the graphene flake. The atomic displacements range from a minimal value of about 0.3~\AA\ to approximately 1.9~\AA\ within the slab, while the largest deviations of up to 2.4~\AA\ occur at the oxygen‑containing functional groups of the flake. The electronic structures (Figure~\ref{fig:FigSTbG}c) display a similar overall character of the total and projected densities of states on the atoms, with the valence band dominated by I and Pb states. The dashed horizontal line indicates the Fermi level, which is slightly shifted in the DFTB/PTBP calculation compared to the DFT/PBEsol result. Overall, the general features of the DOS are well reproduced by DFTB/PTBP compared to DFT/PBEsol.

\subsubsection{Additional ML-MD/TFMC Calculations}\label{sec:MDMC}

\begin{figure}[h!tb] 
	\centering
	\includegraphics[width=1.0\textwidth]{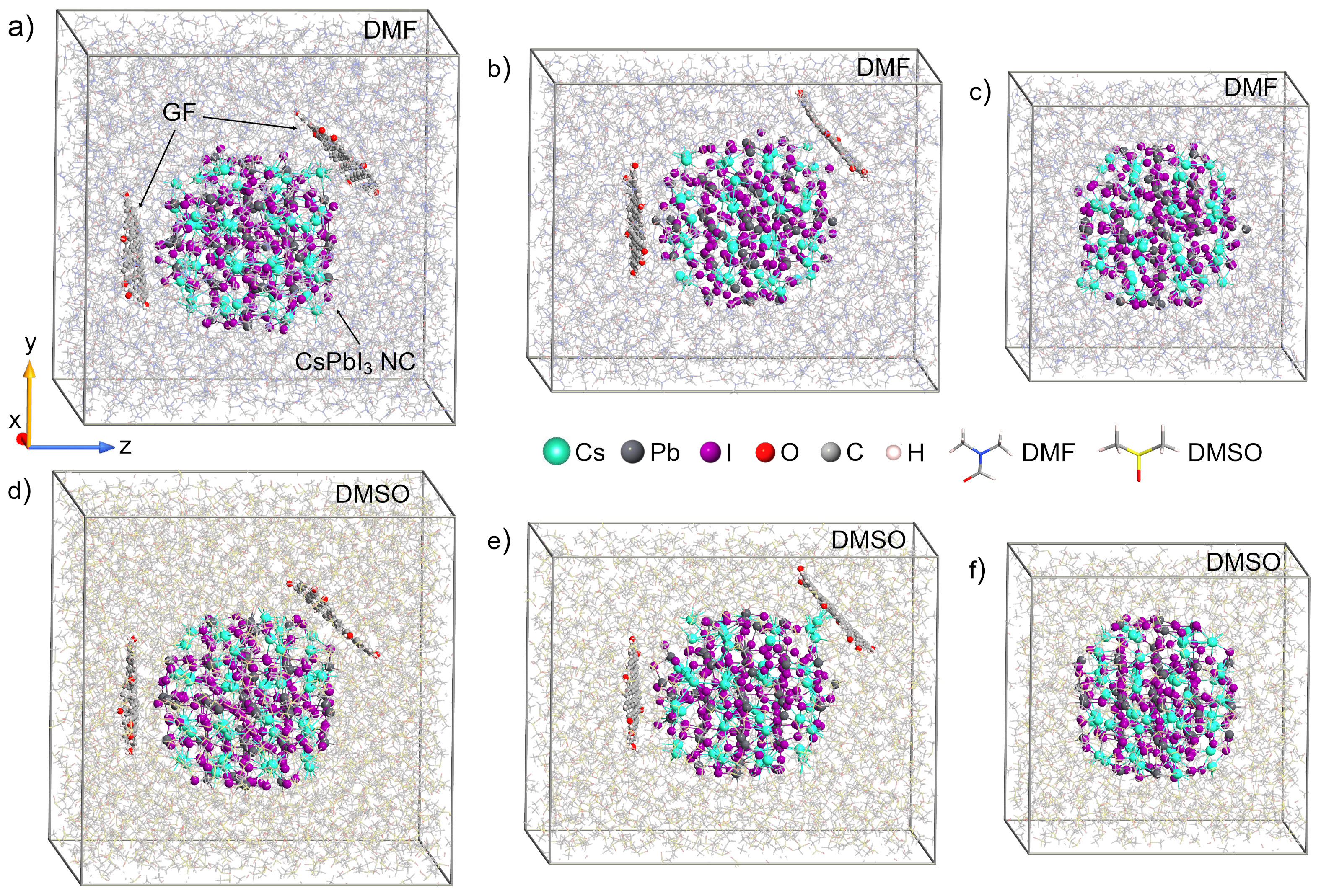} 
	\caption{Snapshots from ML-MD/TFMC calculations of $\mathrm{CsPbI_3}$ nanocrystals (NCs) in DMF and DMSO, with and without two graphene flakes (GF) initially placed above different NC facets. Panels (a,d) show the final configurations of the large simulation box (NC$^\mathrm{bbox}$), containing an NC interacting with two GF in DMF and DMSO, respectively. Panels (b,e) and (c,f) show the corresponding small-box models (NC$^\mathrm{sbox}$) with and without GF, respectively.} 
	\label{fig:s13}
\end{figure}

\begin{figure}[h!tb] 
	\centering
	\includegraphics[width=0.92\textwidth]{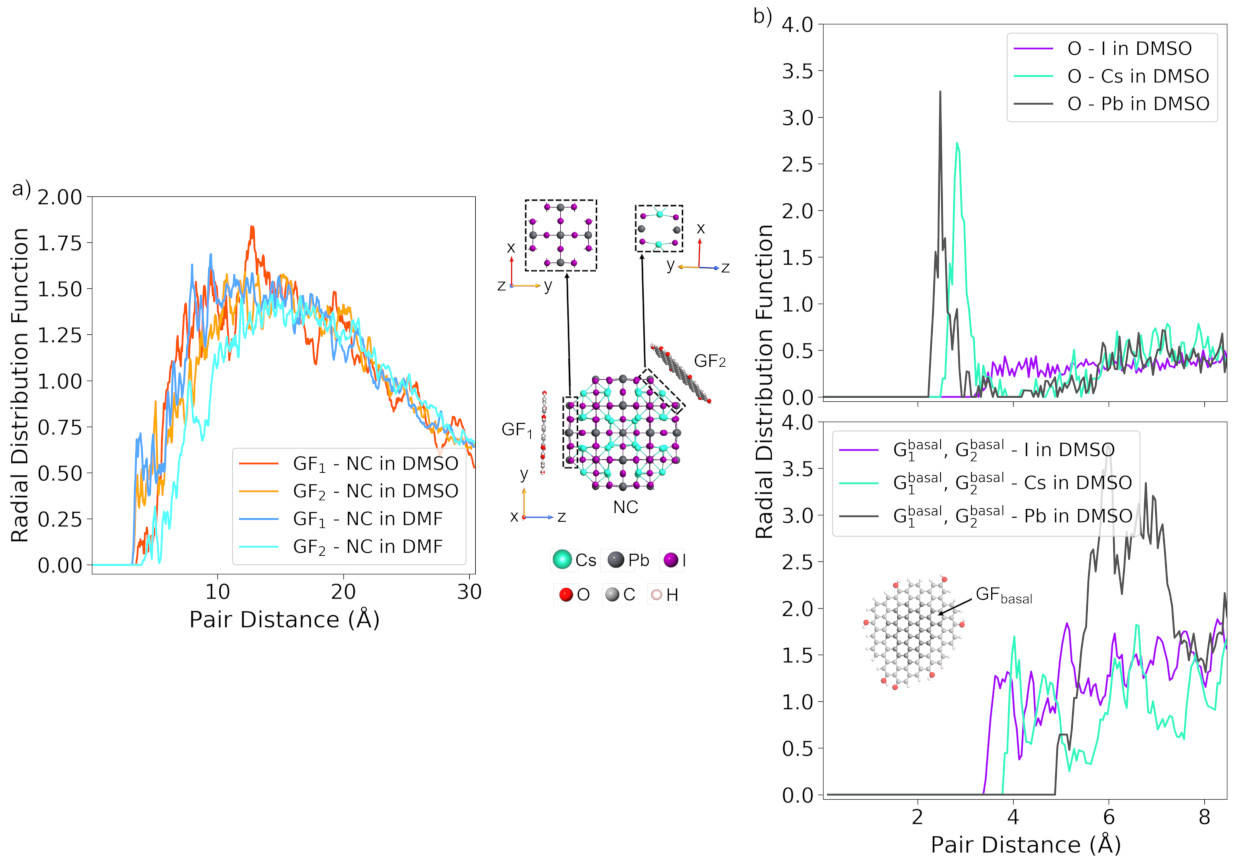} 
	\caption{a) Rolling-averaged radial distribution functions (RDFs) between each GF and the $\mathrm{CsPbI_3}$ NC from the large-box (NC$_\text{GF}^\text{bbox}$ in DMSO and DMF) ML-MD/TFMC simulations in DMSO and DMF. The snapshot on the right shows the initial NC/GF configuration, with GF$_1$ and GF$_2$ placed above two different NC facets. b) RDFs from the large-box DMSO simulation (NC$_\text{GF}^\text{bbox}$ in DMSO). The upper panel shows RDFs between flake O atoms and NC atoms, while the lower panel shows rolling-averaged RDFs between basal-plane C atoms of the two GF and NC atoms. The inset highlights the GF basal-plane atoms used for the analysis, whereas the other flake atoms are shown with higher transparency. RDFs in panels a and b were calculated from the final 1~ps of the large-box simulations and smoothed using a rolling average to reduce noise from limited sampling. In this procedure, each RDF value is replaced by the average of neighbouring points within a 0.25~\AA\ distance window.  
	}
	\label{fig:s14}
\end{figure}

Additional molecular dynamics and time-stamped force-bias Monte Carlo (ML-MD/TFMC) simulations were performed to examine the interaction between $\mathrm{CsPbI_3}$ nanocrystals (NCs) and graphene flakes in DMF and DMSO solvents, two solvents widely used in metal-halide perovskite precursor processing~\cite{BautistaQuijano2023,Wang2018}. The starting NC structure was derived from the simplified $\mathrm{CsPbBr_3}$ nanocrystal model reported by Chen et al.~\cite{Chen2018}, by replacing all Br atoms with I. Two GF were initially placed above two different NC facets: a
PbI$_2$-terminated (100) facet and a Cs/I-rich (101) facet (Figure~\ref{fig:s14}a). To separate solvent effects from GF-induced structural changes, reference systems containing only the $\mathrm{CsPbI_3}$ NC in solvent were also prepared. 
The NC/GF systems were prepared using two box sizes to enable complementary comparisons. Large boxes of $60\times60\times60$~\AA$^3$ were used to compare the final NC/GF configurations in DMF and DMSO under the same box size and sampling time (NC$_{GF}^\text{bbox}$). Smaller NC/GF boxes of $47\times52\times65$~\AA$^3$ (NC$_{GF}^\text{sbox}$), together with NC-only reference boxes of $48\times48\times48$~\AA$^3$ (NC$^\text{sbox}$), were used to assess GF-induced structural changes relative to the corresponding solvent-only NC models (NC$^\text{sbox}$). All boxes were filled with either DMF or DMSO using Packmol~\cite{Martinez2009}. The final configurations are shown in Figure~\ref{fig:s13}.

The simulations were performed in QuantumATK~\cite{Smidstrup2019} using a Osaka PBE-D3 universal machine-learning interatomic potential ($\mathrm{TorchX\_MACE\_Osaka24\_small}$) with a D3(BJ) dispersion correction using PBE damping parameters and a 15~\AA\ cutoff. The Osaka PBE-D3 model is based on the MACE architecture~\cite{batatia2023} and was trained as a multi-domain machine-learning interatomic potential for molecular and crystalline systems~\cite{shiota2025}.
Each system was first propagated by NVE Velocity Verlet ML-MD. Initial velocities were assigned according to a Maxwell--Boltzmann distribution at 300~K. A time step of 0.02~fs was used for systems containing DMF, whereas a time step of 0.1~fs was used for systems containing DMSO. The total ML-MD simulation time was 2.5~ps. The final ML-MD configurations were then used as starting points for TFMC simulations in the NVT ensemble at 300~K. TFMC simulations were performed with a maximum atomic displacement of 0.1~\AA. The TFMC sampling times were 9.7~ps for the small DMF boxes, 10.4~ps for the small DMSO boxes, and 18~ps for the large boxes. All remaining TFMC parameters were kept at their default QuantumATK values. 

Structural analysis of Pb--I, Cs--I, and Cs--Pb bond lengths in the NCs was performed using the final ML-MD/TFMC configurations (Table~\ref{tab:perovskite_bond_lengths}). We additionally analysed local structural trends by calculating radial distribution functions
(RDFs) between selected components of the simulated systems (Figure~\ref{fig:s14}).
For two atom sets, \(A\) and \(B\), the RDF
\(g_{AB}(r)\) describes the probability of finding atoms of set \(B\)
at a distance \(r\) from atoms of set \(A\), relative to a uniform
distribution:
\begin{equation}
	g(r)=\frac{1}{4\pi r^{2}}\frac{1}{\rho N_{A} N_{B}}\sum_{i\in A}\sum_{j\in B, i\neq j}\left \langle \delta (r-\left | r_i -r_j \right |) \right \rangle ,
\end{equation}
where \(N_A\) is the number of atoms in set \(A\), \(\rho_B=N_B/V\) is the number density of atoms in set \(B\), \(\mathbf{r}_i\) and \(\mathbf{r}_j\) are atomic positions, and $\rho$ is the system density.  The RDF profiles for the large-box simulations (NC$_{GF}^\text{bbox}$) were calculated from the final 1~ps of the simulations, whereas those for the small-box (NC$_{GF}^\text{sbox}$, NC$^\text{sbox}$) simulations were calculated from the final 4~ps.

\FloatBarrier
\newpage
\subsection{Experimental Details}
\subsubsection{Materials}

A few layered pristine graphene flakes G3 (Levidian Nanosystems Limited) were used as a carbon additive. According to the manufacturer’s technical specification~\cite{LevidianG3TDS} and product
description~\cite{LevidianG3Product}, G3 is a non-functionalised few-layer graphene nanoplatelet powder exhibiting a combination of flat, wrinkled, and crumpled morphologies. The manufacturer reports a carbon content of $>99.9\%$, oxygen content $<0.1\%$, $sp^2$-carbon bonding content $>99\%$, and a specific surface area of $120\pm20~\mathrm{m^2\,g^{-1}}$. The reported Raman intensity ratios are $I_\mathrm{D}/I_\mathrm{G}<0.8$ and $I_{2\mathrm{D}}/I_\mathrm{G}>0.9$, and no additional impurities were detected by XPS.

Lead iodide (PbI$_2$, 99.999\%), lead bromide (PbBr$_2$, 99.9\%), tin iodide (SnI$_2$, 99.999\%), tin fluoride (SnF$_2$, 99.999\%), fullerene (C$_{60}$, 99.9\%), bathocuproine (BCP, 99.9\%), ammonium thiocyanate (NH4SCN, 99.99\%), glycine hydrochloride (GlyHCl, 99\%), Cesium iodide (CsI, 99.999\%) and F-doped tin oxide (FTO) were purchased from Libra Technology Corporation. Formamidinium iodide (FAI, 99.99\%) and methylammonium iodide (MAI, 99.99\%) were purchased from GreatCell Solar (Australia). (4-(7H-benzo[c]carbazol-7-yl)butyl) phosphonic acid (BCZ, >98.0\%) was purchased from Zhengzhou Alfa Chemical Co., Ltd. Chlorobenzene (CB, 99.8\%, SuperDry, with molecular sieves), N,N-dimethylformamide (DMF, 99.8\%, SuperDry, with molecular sieves), dimethyl sulfoxide (DMSO, 99.7\%, SuperDry, with molecular sieves), ethanol and isopropanol (99.5\%, SuperDry, with molecular sieves) were purchased from J\&K scientific.

\subsubsection{Methods}

\paragraph{Preparation of Graphene Dispersions:}
To evaluate the suitability of different solvents for dispersing a few layered GF, ~1 mg of GF was added to 2~ml of each solvent and subjected to bath sonication for 30~min at ambient temperature. This GF-saturated solution was diluted accordingly to reach the desired relative concentrations. Based on dispersion quality and stability three solvents were chosen for SEM analysis - CHCl$_3$, DMSO, DMF.

\paragraph{Sample preparation for SEM Analysis:}
The samples were prepared by drop-casting the solutions onto flat Si substrates. The coated substrates were dried on a hot plate in a nitrogen-filled glovebox and kept under inert conditions until SEM analysis. The morphology of the dried samples was examined using a Quanta 250F scanning electron microscope operated in high-vacuum mode at an accelerating voltage of 5~kV.

\paragraph{Lead perovskite precursor solution preparation:}
All of the solution preparation steps, except sonication, were performed in a nitrogen-filled glovebox to minimize moisture and oxygen exposure. For drop-casted CsPbI$_3$ precursor solutions, PbI$_2$ (5.9~mg) and CsI (3.3~ mg) were dissolved in 2\,ml either of pure solvent (control sample) or solvent with dispersed graphene. Drop-casted MAPbI$_3$ solutions were prepared according to the same procedure, using PbI$_2$ (5.9~mg) and MAI (2.0~mg) as precursors in the selected solvent systems. In both cases, the solutions were stirred magnetically for 30~min at room temperature. To prepare the Pb perovskite precursor solution for PSCs, 233~mg of FAI, 677.05~mg of PbI$_2$, 8.14 mg of MABr, 28.07~mg of PbBr$_2$, and 19.5 mg of CsI were dissolved in 1~mL of a solvent mixture containing DMF and DMSO (4:1, v/v), and stirred for 8~h at 60~$^o$. The solution was filtered through a 0.22-$\upmu$m hydrophobic PTFE filter before use. For the GF-containing solution, all perovskite precursors were dissolved in GF-containing solvent mixture.

\paragraph{Lead PSCs preparation:}
The ITO glass substrates were sequentially cleaned by ultrasonication in deionized water, acetone, and ethanol, followed by a UV-ozone treatment for 30~min. To deposit the HTL, a solution of BCZ in absolute ethanol (0.5~mg~mL$^{-1}$) was spin-coated onto the ITO substrates at 3000~rpm for 30~s and annealed at 100~$^o$C for 10~min. The perovskite precursor solution was stirred at 60~$^o$C for 6~h and filtered through a 0.22~$\upmu$m polytetrafluoroethylene (PTFE) membrane prior to use. To form the perovskite layer, 100~$\upmu$L of the precursor solution was spin-coated onto the HTL at 5000~rpm for 40~s (acceleration: 1000~rpm s$^{-1}$). At 10~s prior to the end of the spinning process, 200~$\upmu$L of chlorobenzene was dripped onto the spinning film as an antisolvent. The substrates were then annealed at 100~$^o$C for 1~h. Finally, the samples were transferred to a vacuum thermal evaporator. Under a base pressure of <10$^{-5}$ Pa, a 40~nm C$_{60}$ layer (0.03-0.05~nm~s$^{-1}$), a 10~nm BCP layer (0.02-0.03~nm~s$^{-1}$), and a 110~nm Ag electrode (0.1-0.2~nm~s$^{-1}$) were sequentially deposited through a shadow mask defining an active area of 0.0982~cm$^2$.

\paragraph{Tin-lead perovskite precursor solution preparation:}
Previous preparation: 46.8~mg CsI, 185.7~mg FAI, 85.8~mg MAI, 14.1~mg SnF$_2$, 414.9~mg PbI$_2$, 335.3~mg SnI$_2$, 2.7~mg NH$_4$SCN and 4.0~mg GlyHCl were dissolved in 1~mL mixed solvents of DMF:DMSO (3:1) to stir for 8~hours at room temperature. FTO was ultrasonically cleaned in turn by deionized water, acetone, and ethanol for 15~minutes. For the GF-containing target solution, all perovskite precursors were dissolved in GF-containing solvent mixture.

\paragraph{Tin-lead PSCs preparation:}
The devices were fabricated with an architecture of glass/FTO/ PEDOT:PSS/perovskite (with or without Gr)/C$_{60}$/BCP/Ag. First, FTO glass substrates were sequentially cleaned by ultrasonication in deionized water, acetone, and ethanol for 15~min each. After drying, the substrates were treated with UV-ozone for 30~min. The PEDOT:PSS hole-transport layer (HTL) was prepared from an undiluted aqueous dispersion, filtered through a 0.45~$\upmu$m PVDF filter, spin-coated onto the FTO substrates at 4000~rpm for 40~s (acceleration: 2000~rpm~s$^{-1}$), and annealed on a hotplate at 140~$^o$C for 15~min. Subsequently, the substrates were transferred into an N$_2$-filled glovebox. Next, 100~$\upmu$µL of the perovskite precursor solution was spin-coated onto the PEDOT:PSS layer using a two-step process: 1000~rpm for 10~s (acceleration: 600~rpm s$^{-1}$) followed by 4000~rpm for 40~s (acceleration: 2000~rpm~s$^{-1}$). During the second step, an antisolvent chlorobenzene was dripped onto the spinning substrate 20~s prior to the end of the program. The films were then annealed at 100~$^o$ºC for 10~min. Finally, the samples were transferred to a vacuum thermal evaporator. Under a base pressure of <10$^{-5}$~Pa, a 40~nm C$_{60}$ layer (0.03-0.05~nm~s$^{-1}$), a 10~nm BCP layer (0.02-0.03~nm~s$^{-1}$), and a 110~nm Ag electrode (0.1-0.2~nm~s$^{-1}$) were sequentially deposited through a shadow mask defining an active area of 0.0982~cm$^2$.

\paragraph {Device fabrication:}
A total of 92 PSCs were fabricated across eight independent batches, including 48 Pb-based PSCs and 44 Sn–Pb PSCs. For each of the two Pb-based compositions, two independent batches were fabricated under four experimental conditions, with three PSCs prepared for each condition in each batch, giving 24 PSCs per composition. For the Sn–Pb system, four experimental conditions were investigated in four independent batches, giving 11 PSCs per condition and 44 PSCs in total. These numbers refer to the PSCs and do not represent the total number of individual device pixels.

\paragraph{Device Characterization:}
Surface and cross morphology was performed with a scanning electron microscope (Nova NanoSEM 450). Crystalline properties of Sn-Pb perovskite were characterized by X-ray diffractometer (XRD, Bruker D8-ADVANCE). External quantum efficiency (EQE) spectra were measured using a QE-R EQE system (EnLi Technology, Taiwan). 

The J-V measurements were performed using a Keithley 2400 source meter under simulated AM1.5G illumination at 100 mW cm$^{-2}$ provided by an EnliTech SS-X50 solar simulator. The light intensity was calibrated using an EnliTech silicon reference cell certified by NIST and equipped with a KG2 filter.

For the Pb-based devices, the forward scan was performed from -0.1 to 1.2~V, and the reverse scan was performed from 1.2 to -0.1~V. For the Sn-Pb devices, the forward scan was performed from -0.1 to 0.9~V, and the reverse scan was performed from 0.9 to -0.1~V. The scan rate was 20~mV~s$^{-1}$, with a dwell time of 10 ms. The measurements were performed at room temperature in an N$_2$-filled glovebox without preconditioning.

The stabilized power output was measured for 300~s at fixed bias voltages of 0.68~V for the control device and 0.72~V for the Gr-containing device, using an aperture area of 0.0982~cm$^2$. No active maximum-power-point tracking was applied.

The hysteresis index was calculated using HI = (PCE$_\mathrm{max}$ - PCE$_\mathrm{min}$)/PCE$_\mathrm{max}$. Based on the maximum and minimum PCE values obtained from the forward and reverse scans, the hysteresis indices were 0.037 for the control Sn-Pb device and 0.018 for the Gr-containing Sn–Pb device.

\FloatBarrier

\clearpage
\newpage
\section{Additional Modelling Results}
\begin{figure}[h!tb] 
	\centering
	\includegraphics[width=0.85\textwidth]{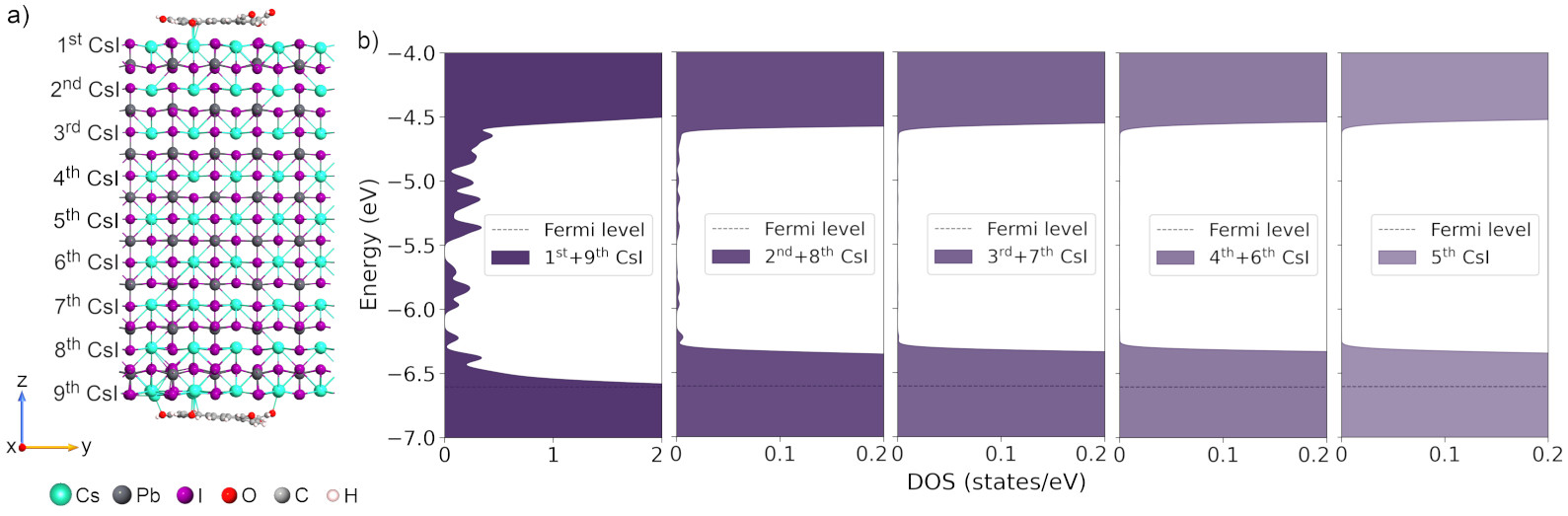} 
	\caption{
		a) Side views of the fully optimized $\mathrm{MAPbI_3}$ $(4 \times 4 \times 3)$ and $\mathrm{CsSnBr_3}$ $(5 \times 5 \times 3)$ slabs shown for the pristine structures and for the systems interfaced with a graphene flake. The corresponding binding energies per interface atom are indicated for each configuration. The elemental color code is shown on the left.
		b) Top views of the perovskite slabs displaying the atomic displacement maps for the $\mathrm{MAPbI_3}$/GF and $\mathrm{CsSnBr_3}$/GF heterostructures. The graphene flakes are indicated as grey shapes, while the atomic colors represent the magnitude of the displacement with respect to the pristine reference structures, ranging from blue (small displacement) to red (largest displacement), revealing the structural distortions induced by the graphene flake. Interatomic connections are shown only for distances below the visualization cutoff (d$_{\mathrm{ij}}$) used for bond detection ($d_{ij}=1.1(R_i+R_j))$, where $R_i$ and $R_j$ are the covalent radii of the atoms). The absence of a connecting line therefore reflects distances exceeding this threshold rather than the absence of bonding interactions.
	}
	\label{fig:s10}
\end{figure}

\begin{figure}[h!tb] 
	\centering
	\includegraphics[width=0.75\textwidth]{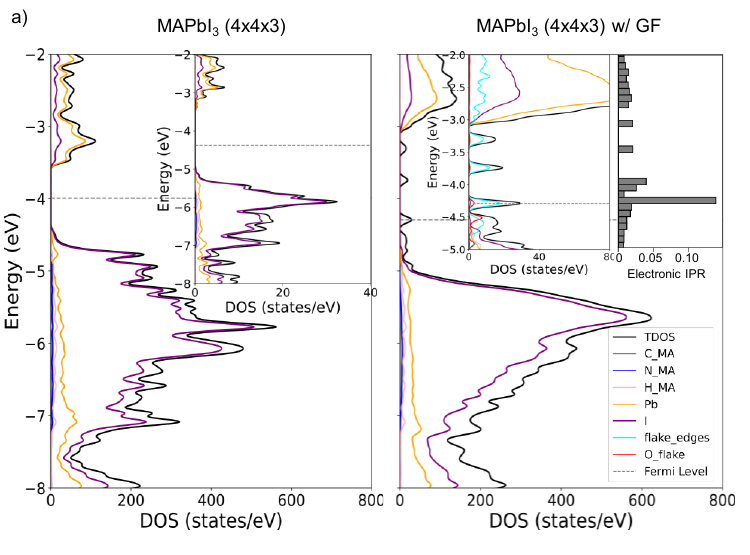} 
	\centering
	\includegraphics[width=0.75\textwidth]{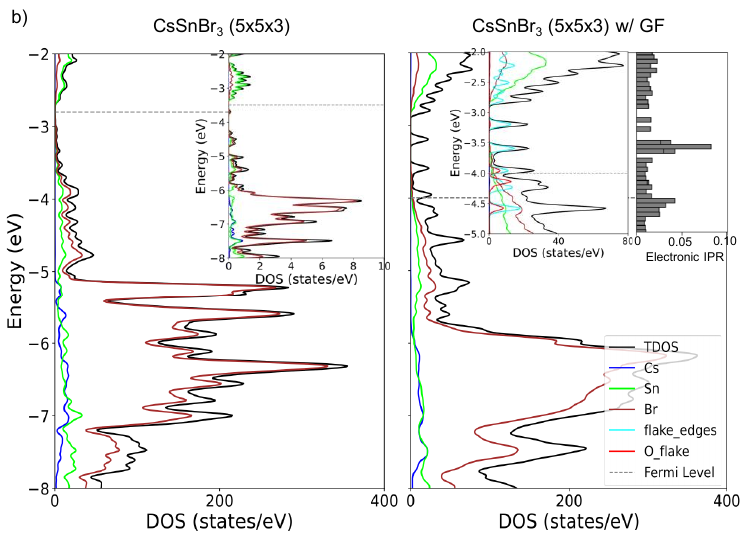} 
	\caption{
		Density of states projected onto selected atoms for a) $\mathrm{MAPbI_3}$ $(4 \times 4 \times 3)$ and b) $\mathrm{CsSnBr_3}$ $(5 \times 5 \times 3)$ slabs, shown for the pristine systems (left) and for the slabs interacting with GF (right). For $\mathrm{MAPbI_3}$, the DOS is projected onto C, N, and H atoms of the MA cations, Pb and I atoms of the perovskite slab, and O atoms and edge C atoms of the graphene flake. For $\mathrm{CsSnBr_3}$, the DOS is projected onto Cs, Sn, and Br atoms of the perovskite slab, and O atoms and edge C atoms of the graphene flake. Insets in the left panels show the DOS projected onto the corresponding atoms for the bulk systems for comparison, while the right panels include enlarged views of the band-gap region together with the corresponding IPR plots.
	}
	\label{fig:s11}
\end{figure}

\begin{figure}[h!tb] 
	\centering
	\includegraphics[width=0.9\textwidth]{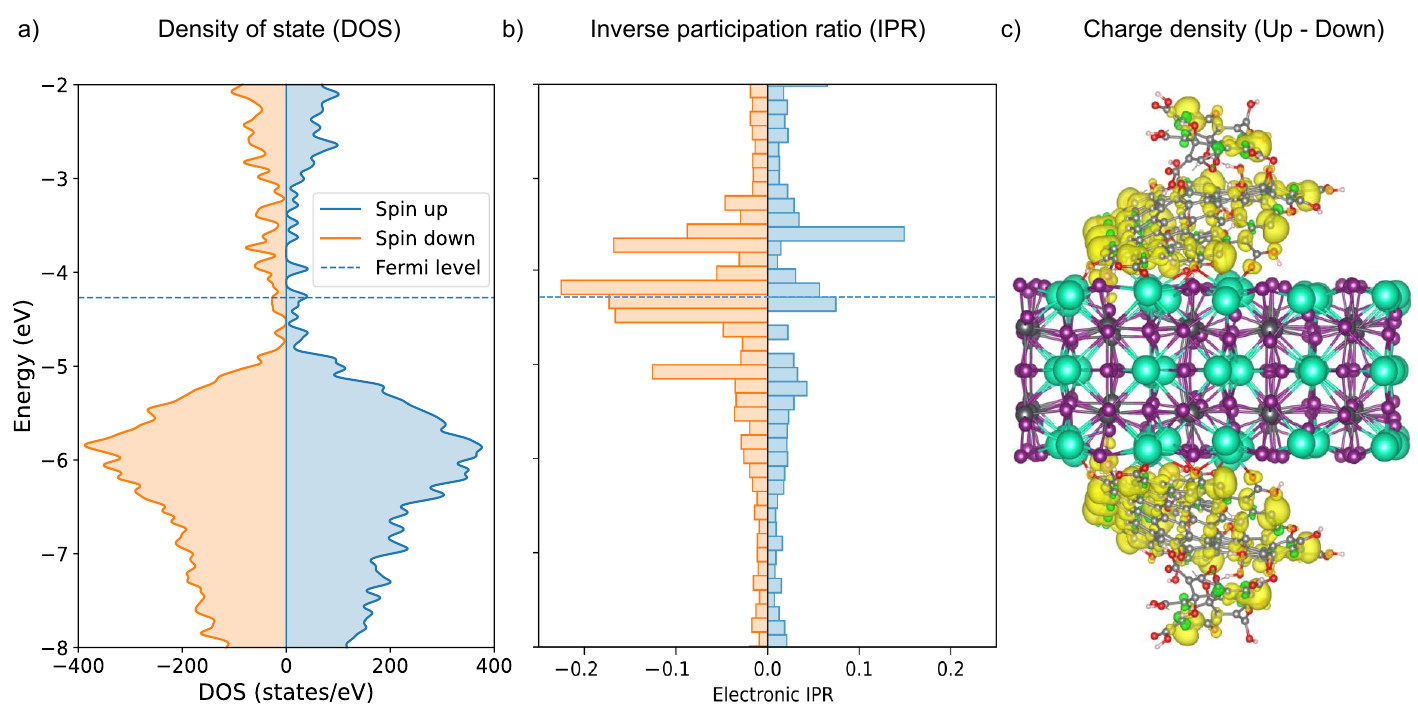} 
	\caption{a) Spin-resolved density of states (DOS) of the $\mathrm{CsPbI_3}$/GOF system, showing contributions from the two spin channels. b)Inverse participation ratio (IPR) for the two spin channels. The dashed horizontal lines in a) and b) indicate the Fermi levels. c) Spin density difference ("Up" - "Down") of the $\mathrm{CsPbI_3}$/GOF. Yellow and green regions indicate an excess of the first ("Up") and second ("Down") spin channel density, respectively. The isosurface value is 0.003 e\,\AA$^{-3}$. The total magnetic moment of the system is 30.7~$\mu_\mathrm{B}$ per simulation cell.
	}
	\label{fig:s8}
\end{figure}

\begin{figure}[h!tb] 
	\centering
	\includegraphics[width=0.9\textwidth]{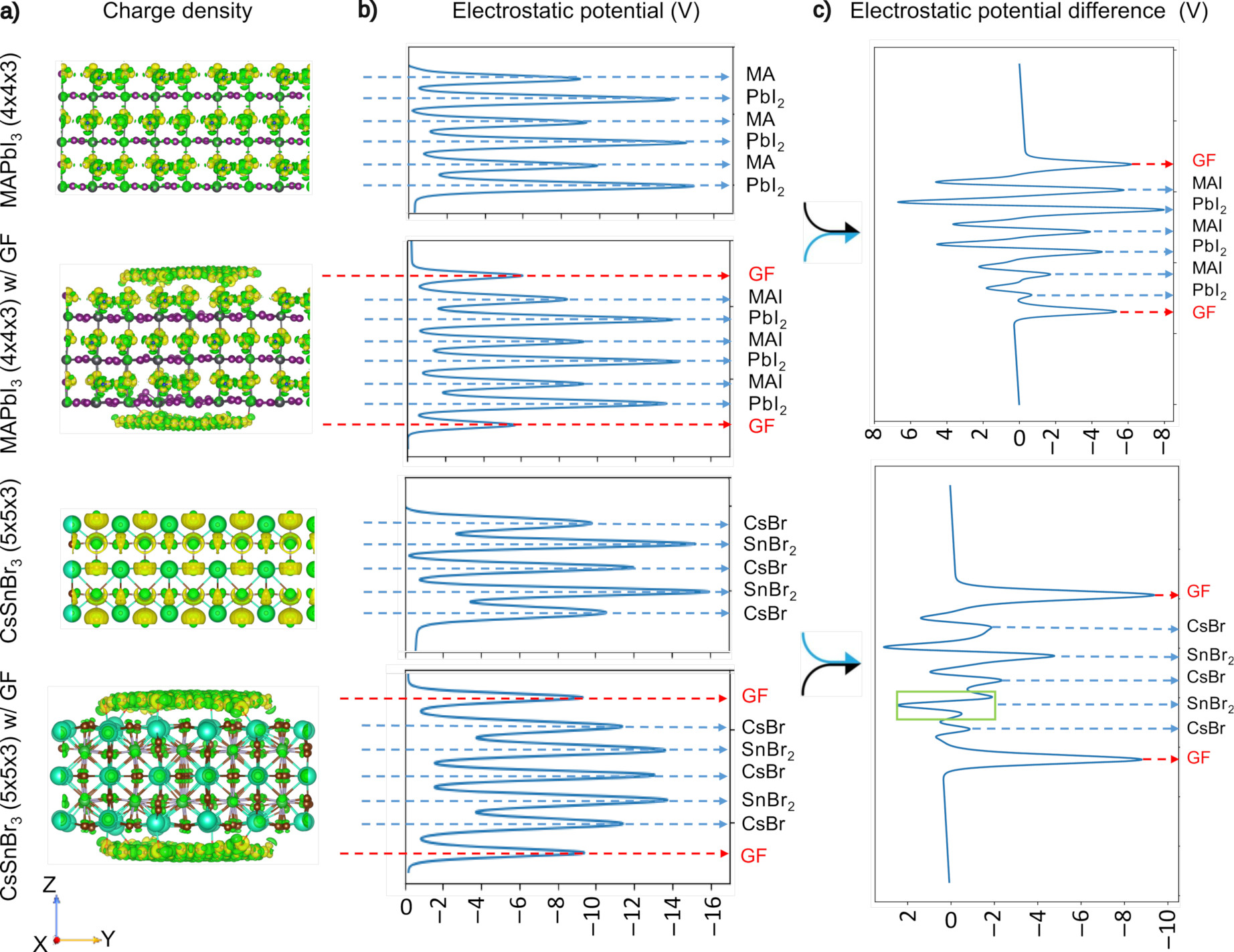} 
	\caption{
		a) Charge-density difference isosurfaces for $\mathrm{MAPbI_3}$ $(4 \times 4 \times 3)$ and $\mathrm{CsSnBr_3}$ $(5 \times 5 \times 3)$ slabs, shown for the pristine systems and for the slabs interacting with graphene flakes (GF). The charge-density difference is defined as the difference between the self-consistent electron density and the superposition of neutral atomic densities. The isovalues are 0.007 and 0.008 e,\AA$^{-3}$ for $\mathrm{MAPbI_3}$ and $\mathrm{MAPbI_3}$/GF, and 0.002 and 0.006 e,\AA$^{-3}$ for $\mathrm{CsSnBr_3}$ and $\mathrm{CsSnBr_3}$/GF, respectively. Green and yellow regions indicate charge accumulation and depletion.
		b) Macroscopically averaged total electrostatic potential profiles along the surface normal ($z$ direction) for pristine $\mathrm{MAPbI_3}$ and $\mathrm{CsSnBr_3}$, and for the corresponding perovskite/GF heterostructures.
		c) Difference in the macroscopically averaged total electrostatic potential along the z-direction for the $\mathrm{MAPbI_3}$/GF and $\mathrm{CsSnBr_3}$/GF systems obtained by subtracting the potential of the corresponding pristine slabs.
	}
	\label{fig:s12}
\end{figure}

\begin{figure}[h!tb] 
	\centering
	\includegraphics[width=0.9\textwidth]{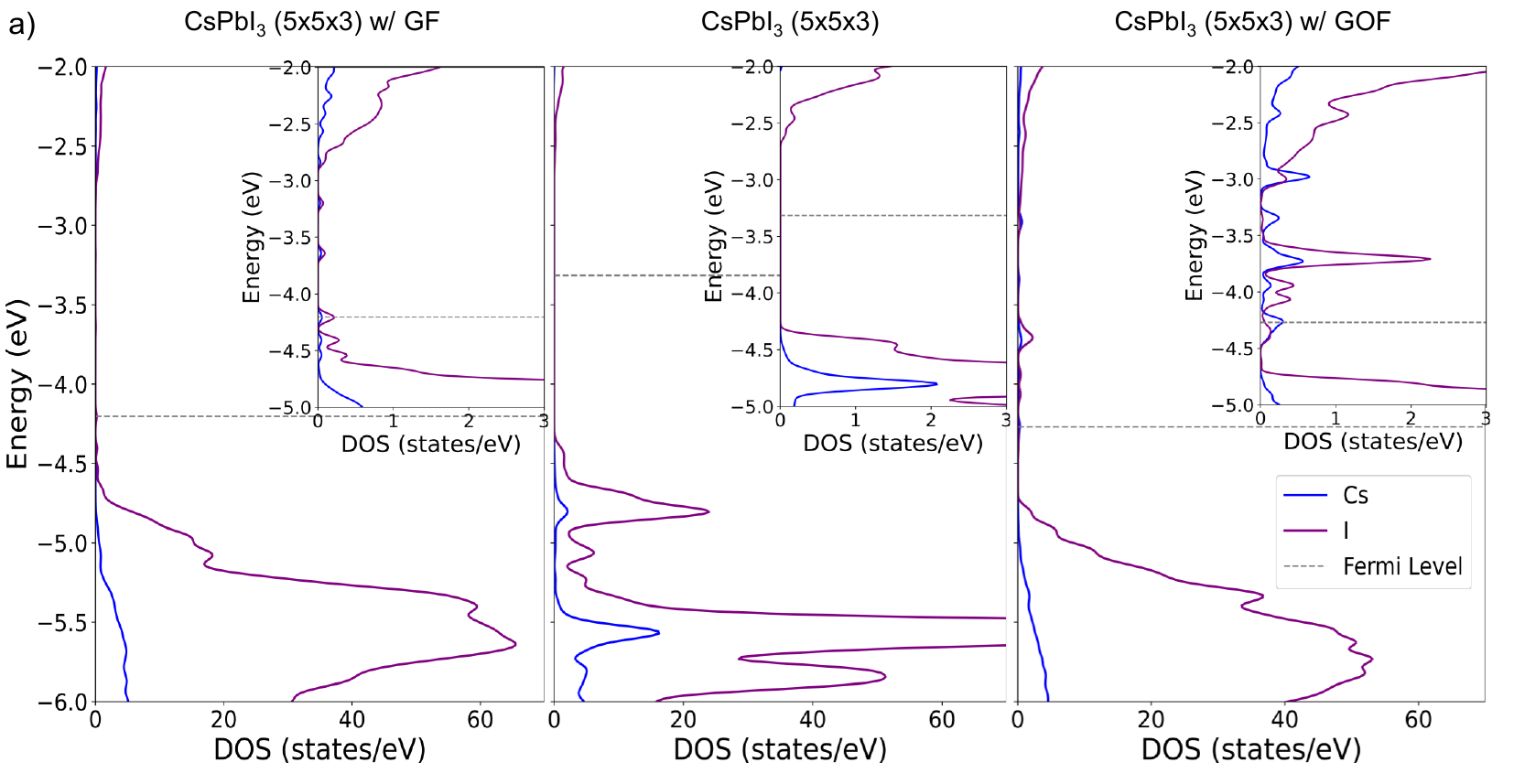} 
	\centering
	\includegraphics[width=0.9\textwidth]{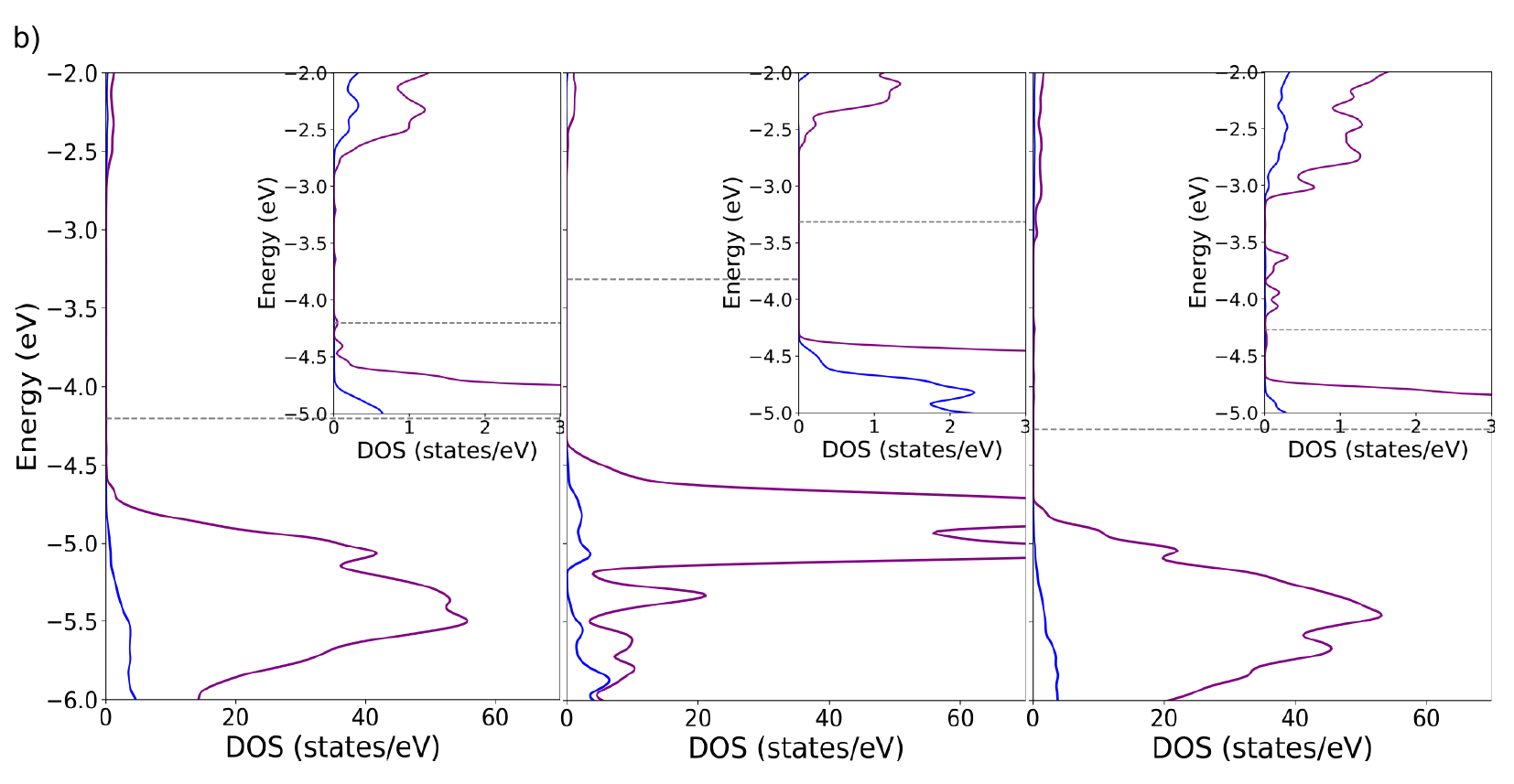} 
	\caption{Density of states (DOS) projected onto Cs and I atoms in a) surface and b) central layers of the $\mathrm{CsPbI_3}$ $(5 \times 5 \times 3)$ slab interacting with GF (left) and GOF (right), compared with the pristine slab (middle). Insets show enlarged views of the band-gap region. The dashed horizontal lines indicates the Fermi levels.}
	\label{fig:s9}
\end{figure}

\begin{figure}[h!tb] 
	\centering
	\includegraphics[width=0.98\textwidth]{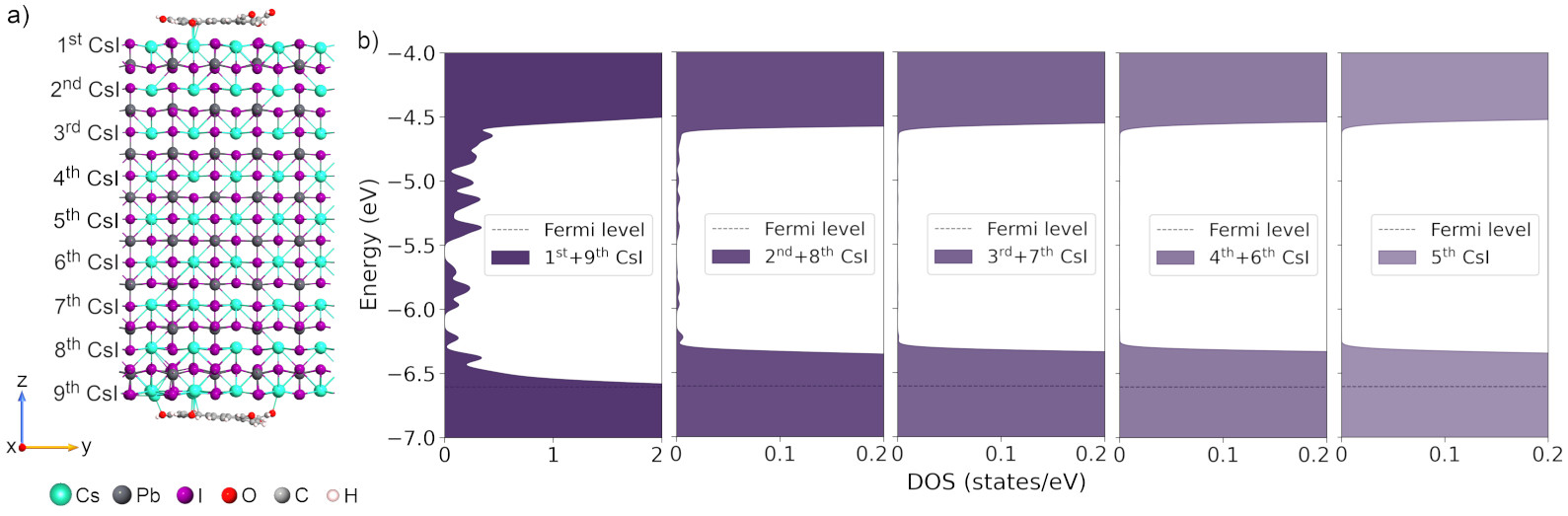} 
	\caption{a) Side view of the fully optimized  $\mathrm{CsPbI_3}$ $(5 \times 5 \times 9)$/GF heterostructure obtained at DFTB/PTBP levels of theory. b) Density of states (DOS) projected onto Cs and I atoms in different CsI layers. }
	\label{fig:s15}
\end{figure}

\newpage
\FloatBarrier
\begin{table}[h!tb]
	\centering
	\caption{Structural, energetic and electronic parameters for all studied bulk and slab perovskite systems with and without the flakes: flake-slab distance ($d_\mathrm{P-F}$), binding energy per atom ($E_{\mathrm{bind}/N}$), 
		Fermi energy ($E_\mathrm{F}$), 
		vacuum-level shift relative to the pristine slab ($\Delta V_{\mathrm{vac}}$), interface dipole ($\upmu/S$), and work-function ($\Phi$). The A Slab refers to the cubic structure with a unit cell ($5\times5\times3$), whereas the B slab refers to the orthorhombic structure with a unit cell ($4\times4\times3$); the subscripts GF and GOF are associated with graphene flakes and graphene oxide flakes, respectively. 
	}
	\begin{tabular}{llccccccc}
		\toprule
		\textbf{} & \textbf{} & $d_\mathrm{P-F}$ (\AA) & $E_{\mathrm{bind}/N}$ (eV) & $E_\mathrm{F}$ (eV) & $\Delta V_{\mathrm{vac}}$ (V) & $\upmu/S$ (C m$^\mathrm{-1}$) & $\Phi$ (eV) \\
		\midrule
		\multirow{4}{*}{$\mathrm{CsPbI_3}$}
		& Bulk & & -3.24 & -3.87  &  &   &  \\
		& A Slab  & &  -3.15 & -3.32 & & & 3.213\\
		& A Slab$_\text{GF}$ & 2.65 -- 3.33 & -5.12 & -4.20 & -0.0090 & -7.97$\times$10$^\mathrm{-14}$ & 4.084 \\
		& A Slab$_\text{GOF}$ & 1.49 -- 1.99  & -5.44 & -4.27& +0.0300 & 2.66$\times$10$^\mathrm{-13}$& 4.193 \\
		\midrule
		\multirow{3}{*}{$\mathrm{MAPbI_3}$}
		& Bulk & & -3.70 & -4.37 & & & \\
		& B Slab &  & -3.67 & -3.99 & & & 4.069 \\
		& B Slab$_\text{GF}$ & 2.56 -- 3.26 & -4.36 & -4.29 & -0.1016 & -8.98$\times$10$^\mathrm{-13}$& 4.268 \\
		\midrule
		\multirow{3}{*}{$\mathrm{CsSnBr_3}$}
		& Bulk & & -3.56 & -3.50 & & &  \\
		& A Slab  & & -3.46 & -3.45 & & & 3.356 \\
		& A Slab$_\text{GF}$  &  2.51 -- 3.37 & -5.29 & -4.00 & -0.0940 & -8.32$\times$10$^\mathrm{-13}$& 3.811 \\
		\bottomrule
	\end{tabular}
	\label{tab:TS2}
\end{table}

\FloatBarrier
\begin{table*}[h!tb]
	\centering
	\caption{Average Pb--I, Cs--I, and Cs--Pb bond lengths in CsPbI$_3$ perovskite slab and nanocrystal systems with and without graphene flake (GF). The bulk and slab models were treated in vacuum and obtained from DFT-optimised geometries. The nanocrystals (NC) were immersed in DMF or DMSO and analysed from the final snapshots obtained after 2.5\,ps MD followed by TFMC simulations of 9.7\,ps in DMF and 10.4\,ps in DMSO for the small simulation boxes, and 18\,ps for the big simulation boxes (Figure~\ref{fig:s13}). The NC$^\text{sbox}$ and NC$^\text{bbox}$ refer to the simulation box size. Values are reported as mean $\pm$ standard deviation in \AA. The bond-length variation $\Delta$d=d$_{\mathrm{GF}}$-d is reported with propagated uncertainty for the corresponding system.
	}
	\begin{tabular}{lllccc}
		\toprule
		\textbf{} & \textbf{} & \textbf{}
		& Pb--I
		& Cs--I
		& Cs--Pb \\
		\midrule
		\multirow{4}{*}{vacuum}
		& Bulk
		& $d$ (\AA)
		& 3.11
		& 4.40
		& 5.38 \\
		& A Slab
		& $d$ (\AA)
		& 3.103$\pm$0.047
		& 4.189$\pm$0.166
		& 4.770$\pm$0.010 \\
		& \multirow{2}{*}{A Slab$_\text{GF}$}
		& $d$ (\AA)
		& 3.139$\pm$0.061
		& 3.964$\pm$0.152
		& 4.594$\pm$0.200 \\
		&
		& $\Delta d$ (\AA)
		& +0.036$\pm$0.077
		& -0.225$\pm$0.225
		& -0.176$\pm$0.200 \\
		\midrule
		\multirow{4}{*}{DMSO}
		& NC$^\text{sbox}$
		& $d$ (\AA)
		& 3.222$\pm$0.228
		& 4.061$\pm$0.580
		& 5.248$\pm$0.699 \\
		& \multirow{2}{*}{NC$_\text{GF}^\text{sbox}$}
		& $d$ (\AA)
		& 3.306$\pm$0.557
		& 3.942$\pm$0.432
		& 5.067$\pm$0.672 \\
		&
		& $\Delta d$ (\AA)
		& +0.084$\pm$0.602
		& -0.119$\pm$0.723
		& -0.181$\pm$0.970 \\
		& NC$_\text{GF}^\text{bbox}$
		& $d$ (\AA)
		& 3.190$\pm$0.149
		& 4.107$\pm$0.419
		& 5.038$\pm$0.753 \\
		\midrule
		\multirow{4}{*}{DMF}
		& NC$^\text{sbox}$
		& $d$ (\AA)
		& 3.237$\pm$0.281
		& 4.033$\pm$0.492
		& 5.252$\pm$0.690 \\
		& \multirow{2}{*}{NC$_\text{GF}^\text{sbox}$}
		& $d$ (\AA)
		& 3.281$\pm$0.372
		& 4.334$\pm$0.954
		& 5.07$\pm$0.818 \\
		&
		& $\Delta d$ (\AA)
		& +0.045$\pm$0.466
		& +0.302$\pm$1.073
		& -0.174$\pm$1.070 \\
		& NC$_\text{GF}^\text{bbox}$
		& $d$ (\AA)
		& 3.221$\pm$0.347
		& 4.030$\pm$0.486
		& 5.003$\pm$0.495 \\
		\bottomrule
	\end{tabular}
	\label{tab:perovskite_bond_lengths}
\end{table*}

\newpage
\FloatBarrier

\section{Additional Experimental Results}

\begin{figure}[h!tb] 
	\centering
	\includegraphics[width=0.6\textwidth]{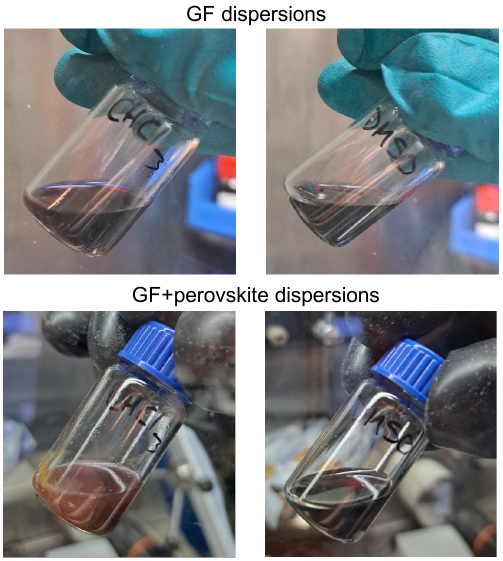} 
	\caption{Pictures of GF and perovskite+GF (0.1\,M) dispersions in CHCl$_3$ and DMSO.}
	\label{fig:FigSE1}
\end{figure}

\begin{figure}[h!tb] 
	\centering
	\includegraphics[width=0.5\textwidth]{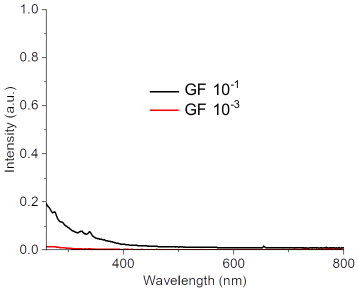} 
	\caption{UV-vis spectra of GF dispersions in DMSO at the minimum and maximum concentrations used.}
	\label{fig:FigSE2}
\end{figure}

\begin{figure}[h!tb]
	\centering
	\includegraphics[width=1.0\textwidth]{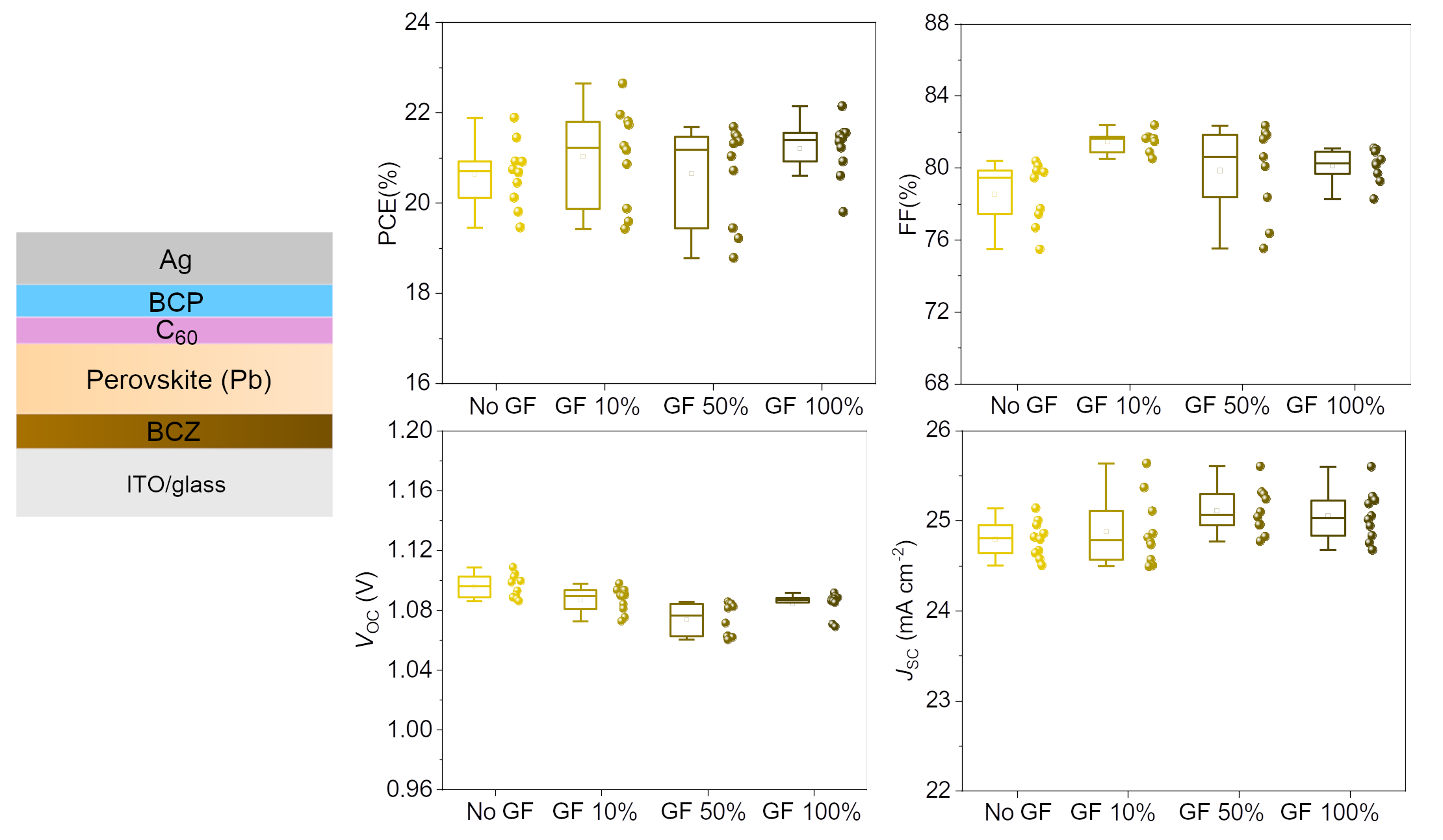} 
	\caption{Distribution of PV parameters of Pb-based perovskite solar cells with Cs$_{0.05}$(FA$_{0.95}$MA$_{0.05}$)$_{0.95}$Pb(I$_{0.95}$Br$_{0.05}$)$_{3}$ perovskite composition and 
		ITO/BCZ/Perovskite/C$_{60}$/BCP/Ag device architecture, with different GF concentrations.}
	\label{fig:FigSE3}
\end{figure}

\begin{figure}[h!tb]
	\centering
	\includegraphics[width=0.5\textwidth]{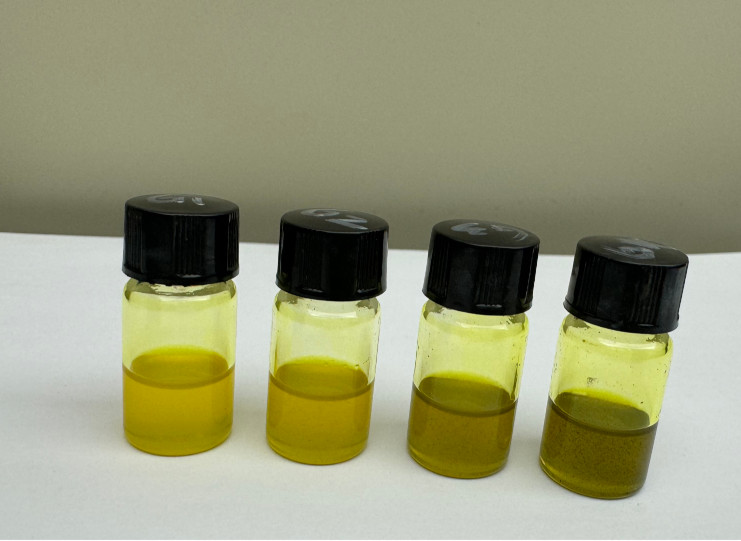} 
	\caption{From left to right, perovskite precursor solution with increasing concentration of GF (0, 10\%, 50\%, 100\%) before solution filtering.}
	\label{fig:FigSE4}
\end{figure}

\begin{figure}[h!tb]
	\centering
	\includegraphics[width=1.0\textwidth]{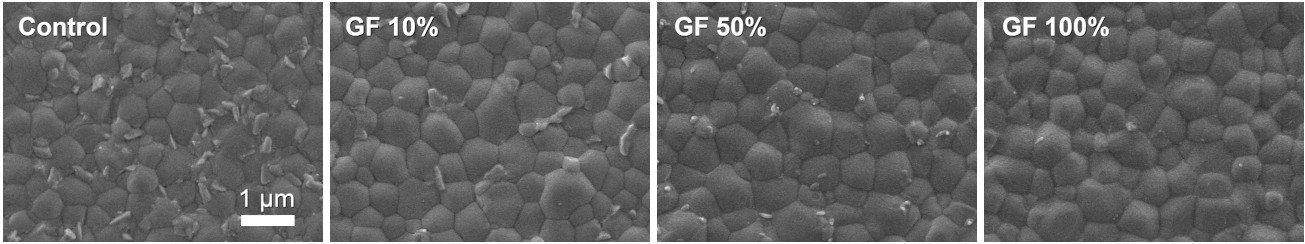} 
	\caption{SEM of spin-coated mixed Sn–Pb perovskite films with increasing concentrations of graphene. 100\% GF is the concentration directly obtained from the filtered 1~wt\% dispersion. The less concentrated dispersions were prepared by simple dilution. }
	\label{fig:FigSE5}
\end{figure}

\begin{figure}[h!tb]
	\centering
	\includegraphics[width=0.85\textwidth]{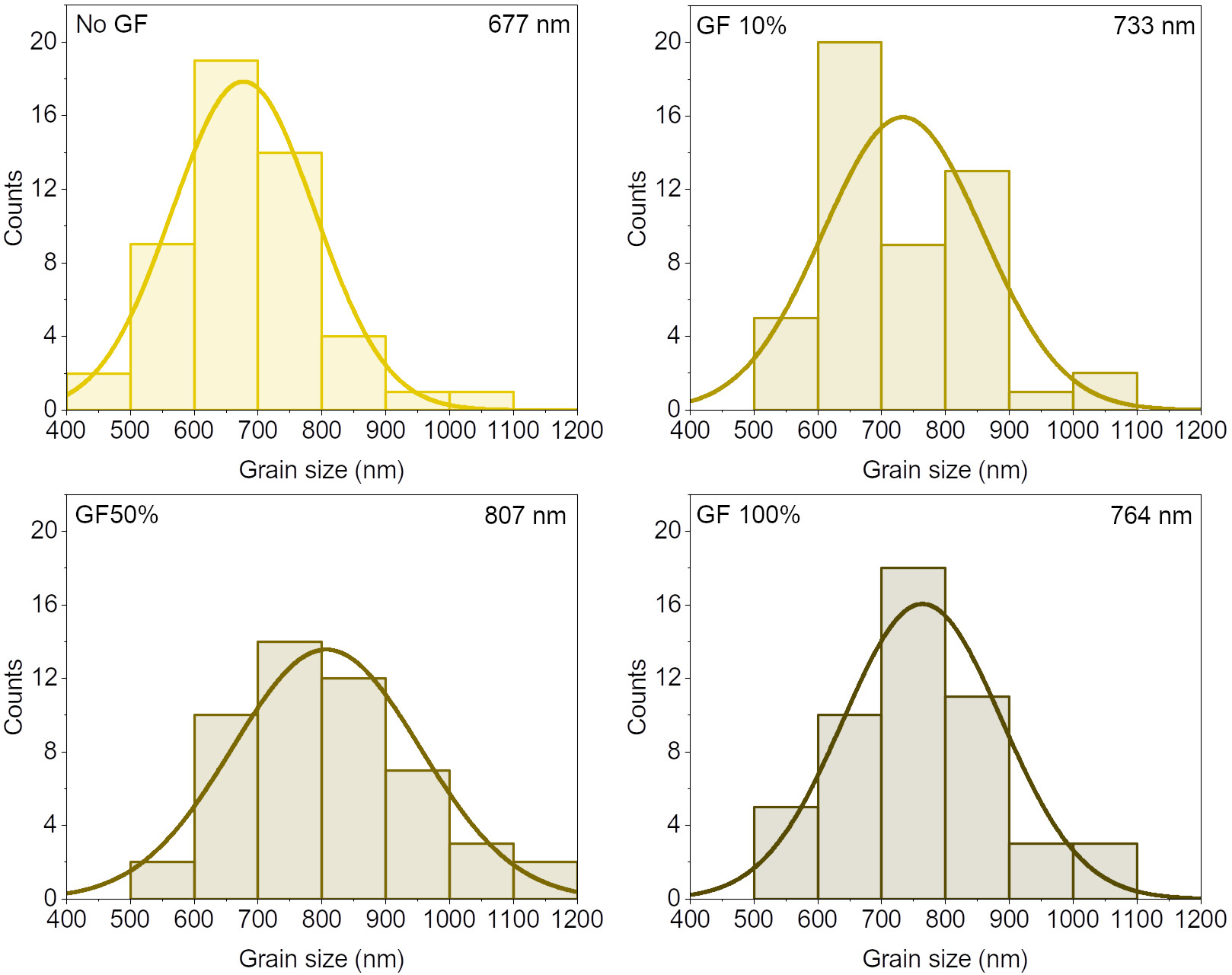} 
	\caption{Grain size distribution for each CsPbI$_3$ thin films at different GF concentrations.}
	\label{fig:FigSE6}
\end{figure}

\begin{figure}[h!tb]
	\centering
	\includegraphics[width=0.6\textwidth]{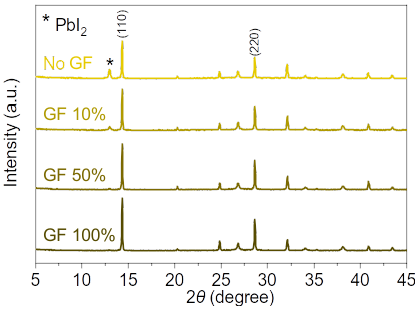} 
	\caption{XRD patterns of Sn--Pb perovskite films with different concentrations of GF.}
	\label{fig:FigSE7}
\end{figure}

\begin{figure}[h!tb]
	\centering
	\includegraphics[width=0.85\textwidth]{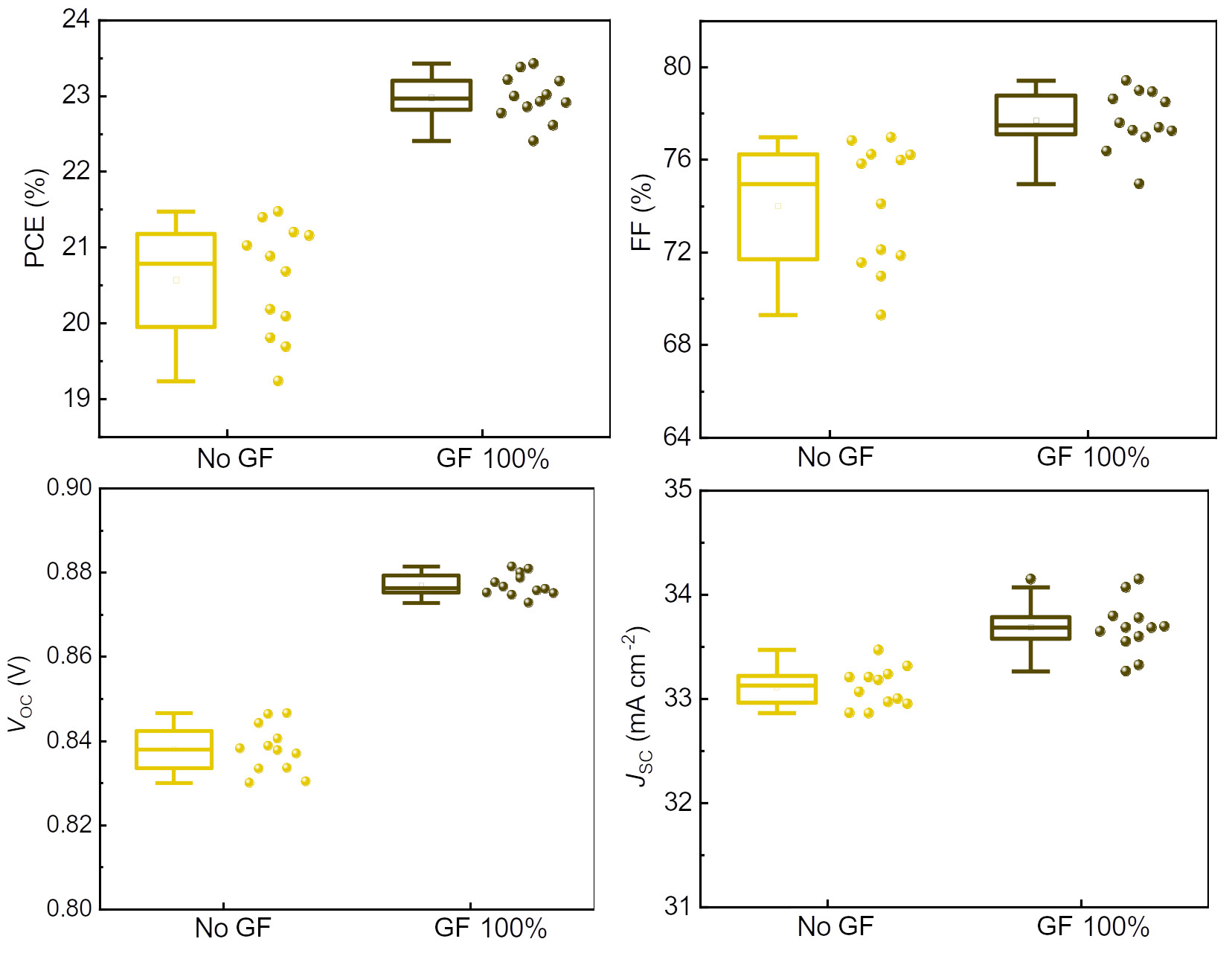} 
	\caption{Distribution of PV parameters values for Sn--Pb devices with and without the optimum concentration of GF.}
	\label{fig:FigSE8}
\end{figure}

\begin{figure}[h!tb]
	\centering
	\includegraphics[width=0.45\textwidth]{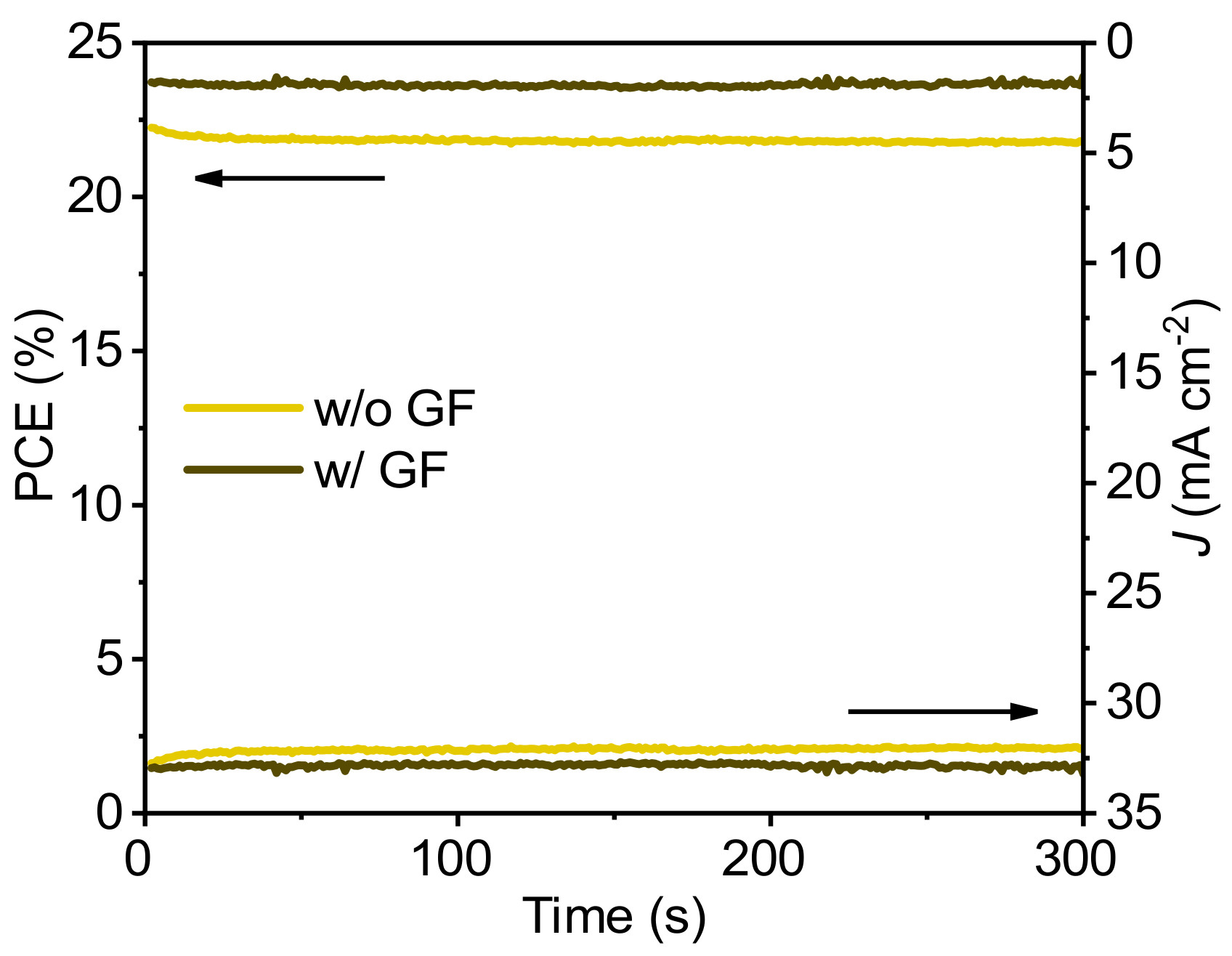} 
	\caption{SPO of mixed Sn--Pb PSCs with and without the optimum concentration of GF.}
	\label{fig:FigSE10}
\end{figure}

\begin{figure}[h!tb]
	\centering
	\includegraphics[width=0.45\textwidth]{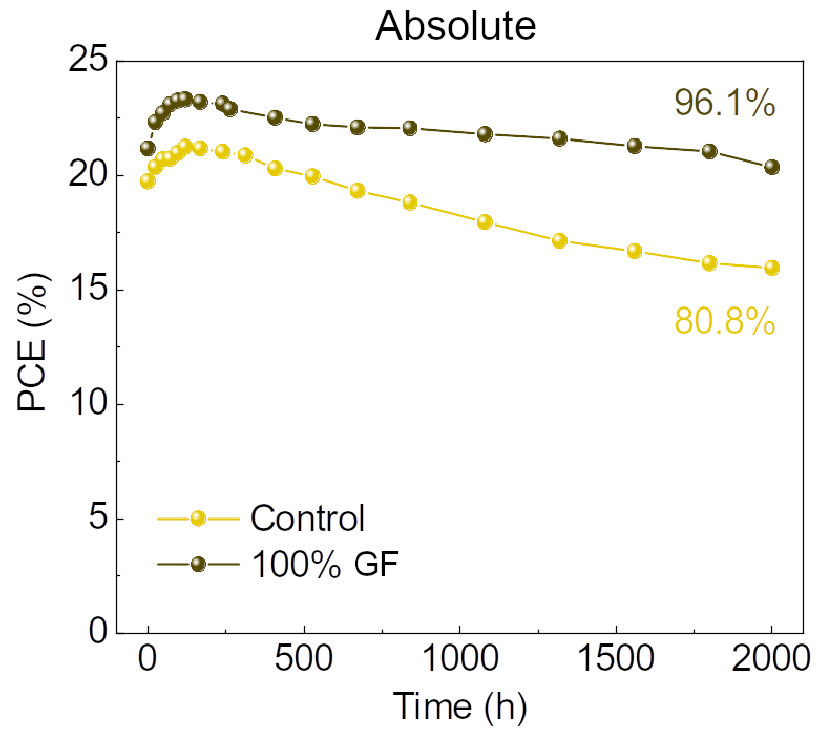} 
	\caption{Evolution of the power conversion efficiency (PCE) of control and 100\% GF-containing perovskite solar cells during ageing.}
	\label{fig:FigSE9}
\end{figure}

\FloatBarrier


\FloatBarrier
\clearpage
\bibliographystyle{rsc}
\bibliography{ref}

\end{document}